\documentclass{elsart}

\usepackage[british]{babel}
\usepackage[utf8]{inputenc}

\usepackage{graphicx,algorithm,algorithmic}
\usepackage{url}
\usepackage{subfigure}
\usepackage{color}
\usepackage{amssymb}
\usepackage{multirow, array}

\begin{document}

\begin{frontmatter}

\title{FavorQueue: a Parameterless Active Queue Management to Improve TCP Traffic Performance}

\author{Pascal Anelli\corauthref{cor1}\thanksref{label1}}
\ead{pascal.anelli@univ-reunion.fr}
\author{R\'emi Diana\thanksref{label2}\thanksref{label3}}
\ead{remi.diana@isae.fr}
\author{Emmanuel Lochin\thanksref{label2}}
\ead{emmanuel.lochin@isae.fr}
\corauth[cor1]{Corresponding author. Address: Universit\'e de la R\'eunion LIM, BP 7151, 2 rue Joseph Wetzel, 97490 Sainte Clotilde, France}
\address[label1]{Universit\'e de la R\'eunion - EA2525 LIM, Sainte Clotilde, France}
\address[label2]{Universit\'e de Toulouse; ISAE; Toulouse, France}
\address[label3]{T\'eSA/CNES/Thales Alenia Space, Toulouse, France}


\maketitle


\begin{abstract}
This paper presents and analyses the implementation of a novel active queue management (AQM) named FavorQueue that aims to improve delay transfer of short lived TCP flows over best-effort networks. 
The idea is to dequeue packets that do not belong to a flow previously enqueued first. The rationale is to mitigate the delay induced by long-lived TCP flows over the pace of short TCP data requests and to prevent dropped packets at the beginning of a connection and during recovery period. 
Although the main target of this AQM is to accelerate short TCP traffic, we show that FavorQueue does not only improve the performance of short TCP traffic but also improves the performance of all TCP traffic in terms of drop ratio and latency whatever the flow size. In particular, we demonstrate that FavorQueue reduces the loss of a retransmitted packet, decreases the number of dropped packets recovered by RTO and improves the latency up to 30\% compared to DropTail. 
Finally, we show that this scheme remains compliant with recent TCP updates such as the increase of the initial slow-start value.
\end{abstract}


\end{frontmatter}

\section{Introduction}
\label{sec:intro}

Internet is still dominated by web traffic running on top of short-lived TCP connections \cite{IOR2009}. Indeed, as shown in \cite{ciullo09}, among 95\% of the client TCP traffic and 70\% of the server TCP traffic have a size smaller than ten packets. This follows a common web design practice that is to keep viewed pages lightweight to improve interactive browsing in terms of response time \cite{chen03}. In other words, the access to a webpage often triggers several short web traffics that allow to keep the downloaded page small and to speed up the display of the text content compared to other heavier components that might compose it\footnote{See for instance: "Best Practices for Speeding Up Your Web Site" from Yahoo developer networki: \url{http://developer.yahoo.com/performance/rules.html}} (\textit{e.g.} pictures, multimedia content, design components). As a matter of fact, following the growth of the web content, we can still expect a large amount of short web traffic in the near future. 

TCP performance suffers significantly in the presence of bursty, non-adaptive cross-traffic or when the congestion window is small (\textit{i.e.} in the slow-start phase or when it operates in the small window regime).  
Indeed, bursty losses, or losses during the small window regime, may cause Retransmission Timeouts (RTO) which trigger a slow-start phase. In the context of short TCP flows, TCP fast retransmit cannot be triggered if not enough packets are in transit. As a result, the loss recovery is mainly done thanks to the TCP RTO and this strongly impacts the delay.
Following this, in this study we seek to improve the performance of this pervasive short TCP traffic without impacting on long-lived TCP flows. We aim to exploit router capabilities to enhance the performance of short TCP flows over a best-effort network, by giving a higher priority to a TCP packet if no other packet belonging to the same flow is already enqueued inside a router queue. The rationale is that isolated losses (for instance losses that occur at the early stage of the connection) have a strong impact on the TCP flow performance than losses inside a large window. Then, we propose an AQM, called FavorQueue (FaQ), which allows to better protect packet retransmission and short TCP traffic when the network is severely congested.

\begin{figure}[htb!]
   \begin{minipage}[b]{1.0\columnwidth}
	\centering
	\includegraphics[width=0.7\columnwidth]{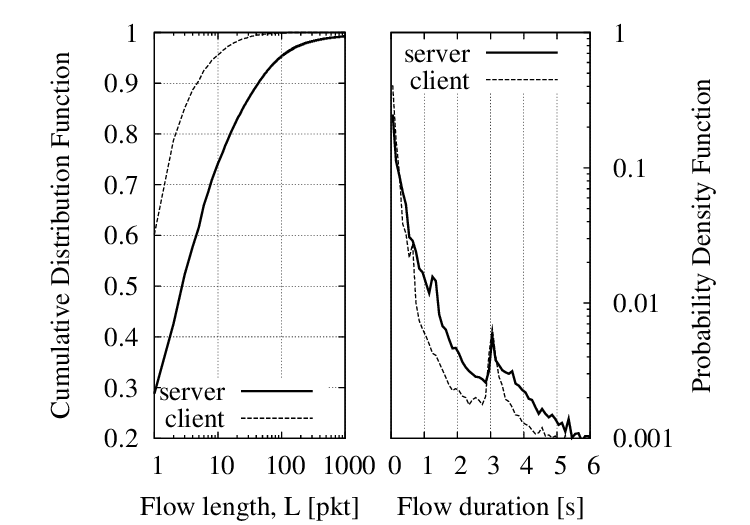}
	\caption{TCP flow length distribution and latency (by courtesy of the authors of \cite{ciullo09}).}
	\label{fig:ciullo09}
   \end{minipage}
   
   \begin{minipage}[b]{1.0\columnwidth}   
	\centering
  	\includegraphics[width=0.7\columnwidth]{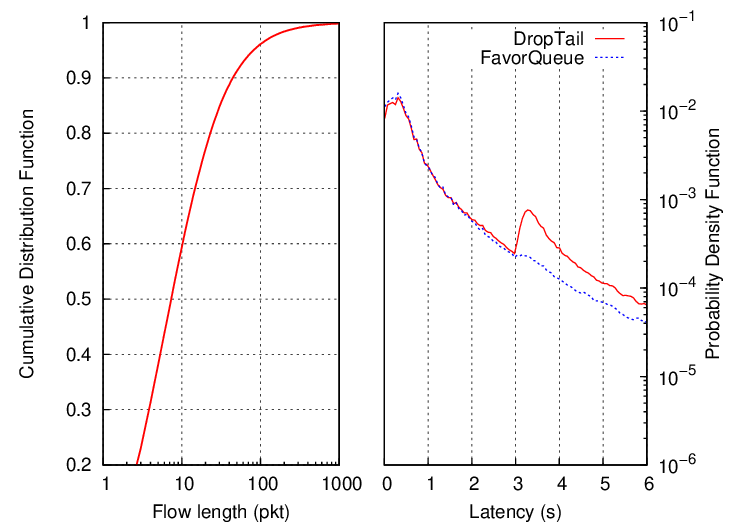}
  	\caption{TCP flow latency distribution from our simulation model.}
  	\label{fig:delaypdf}
   \end{minipage}
\end{figure}

In order to give the reader a clear view of the problem we tackle with our proposal, we rely on paper \cite{ciullo09}. 
Figure \ref{fig:ciullo09} shows that the flow duration (or latency\footnote{The latency refers to the delay between the first packet sent and the last received.}) 
of short TCP traffic is strongly impacted by an initial lost packet which is recovered later by an RTO. 
Indeed, at the early stage of the connection, the number of packets exchanged is too small to allow an accurate RTO estimation. Thus, a retransmission is triggered by the default RTO time value which is set to three seconds \cite{rfc1122}. In this figure, the authors also give the cumulative distribution function (CDF) of TCP flow length and the probability density function of their completion time from an experimental measurement dataset obtained during one day on a ISP BRAS link which aggregates more than 30,000 users. We have simulated a similar experimental scenario with ns-2 (\textit{i.e.} with a similar flow length CDF according to a Pareto distribution) and obtained a similar probability density function of the TCP flows duration as shown in Figure \ref{fig:delaypdf} for the DropTail queue curve. Both figures (\ref{fig:ciullo09} and \ref{fig:delaypdf}) clearly highlight a latency peak at $t=3$ seconds which corresponds to this default RTO value \cite{rfc1122}. In this experiment scenario, 56\% of dropped packets are recovered after an RTO expires (versus 70\% in the experiments of \cite{ciullo09}). As a matter of fact, these experiments show that the success of the TCP slow-start completion is a key performance indicator.
The second curve in Figure \ref{fig:delaypdf}, shows the result we obtain by using our proposal called FavorQueue (FaQ). Clearly, the peak previously emphasized has disappeared. This means the initial losses that strongly impacted the TCP traffic performance have decreased.

An important contribution of this work is the demonstration that our scheme, by favoring isolated TCP packets, decreases the latency by decreasing the loss ratio of short TCP flows without impacting long TCP traffic. 
However, as FavorQueue does not discriminate short from long TCP flows, every flows take advantage of this mechanism when entering either the slow-start or a recovery phase. 
Our evaluations show that 58\% of short TCP flows improve their latency and that 80\% of long-lived TCP flows also take advantage of this AQM. For all sizes of flows, on average, the expected gain of the transfer delay is about 30\%. This gain results from the decrease of the drop ratio of non opportunistic flows which are those that less occupy the queue. Furthermore, the more loaded the queue, the bigger the impact of FavorQueue. Indeed, when there is no congestion, FavorQueue does not have any effect on the traffic. In other words, this proposal is activated only when the network is severely congested.

Finally, FavorQueue does not request any transport protocol modification. Although we talk about giving a priority to certain packets, there is no per-flow state needed inside the FavorQueue router. This mechanism must be seen as an extension of DropTail that greatly enhances TCP sources performance by favoring (more than prioritizing) certain TCP packets. 
We present related work in Section \ref{sec:related} where we position FavorQueue with respect to other propositions.
Then, Section \ref{sec:description} describes the design of the proposed scheme. In Section \ref{sec:exp}, we present the experimental methodology used in this paper. Sections \ref{sec:perf} and \ref{sec:understand} dissects and analyses the performance of FavorQueue. Following these experiments and statistical analysis, we propose a stochastic model of the mechanism Section \ref{sec:model}. We then propose to assess the performance of FavorQueue over a realistic example when only the edge router enables this AQM scheme in Section \ref{sec:5hops}. 
Finally, we propose to discuss the implementation and some security issues in Section \ref{sec:discuss} and conclude this work in Section \ref{sec:conclu}.

\section{Related work}
\label{sec:related}

Several improvements have been proposed in the literature and at the IETF to attempt to solve the problem of short TCP flows performance. Existing solutions can be classified into three different action types: (1)~to enable a scheduling algorithm at the router queue level; (2)~to give a priority to certain TCP packets or (3)~to act at the TCP level in order to decrease the number of RTO or the loss probability. Concerning the two first items, the solution involves the core network while the third one involves modifications at the end-host. In this related work, we first place FaQ among several core network solutions and then explain how FaQ might complete end-hosts' solutions. 

\subsection{Enhancing short TCP flows performance inside the core network}

\subsubsection{The case of short and long TCP flows differentiation}

Several studies \cite{Kantawala02}\cite{avrachenkov04}\cite{rai05} have proposed to serve first short TCP traffic to improve the overall system performance. These studies follow one queueing theory result which stands that the overall mean latency is reduced when the shortest job is served first \cite{Kleinrock75}. 
One of the precursor in the area is \cite{rai05}, where the authors proposed to adapt the Least Attained Service (LAS) \cite{Kleinrock75}, which is a scheduling mechanism that favors short jobs without prior knowledge of job sizes, for packets networks. 
As for FavorQueue, LAS is not only a scheduling discipline but a buffer management mechanism. This mechanism is similar to FavorQueue principle since the priority of the packet is given without knowlegde of the size of the flow and the classification is closely related to the buffer management scheme.  However, the next packet serviced under LAS is the one that belongs to the flow that has received the least amount of service. By this definition, LAS will serve packets from a newly arriving flow until that flow has received an amount of service equal to the amount of least service received by a flow in the system before its arrival. Compared to LAS, FavorQueue has no notion of amount of service as we seek to favor short job by accelerating their connection establishement. Thus, there is no configuration and no complex settings.

In \cite{Kantawala02} and \cite{avrachenkov04}, the authors push the same idea further and attempt to differentiate short from long TCP flows according to a scheduling algorithm. 
The differences between these solutions are based on the number of queues used which are either flow stateless or stateful. 
In \cite{avrachenkov04}, Running Number 2 Class differentiation mechanism (RuN2C) uses an AQM which enables a push out algorithm to protect short TCP flow packets from loss. Short TCP flows identification is done inside the router by looking at the TCP sequence number.
However and in order to correctly distinguish short from long TCP flows, the authors modify the standard TCP sequence numbering which involves a major modification of the TCP/IP stack.
In \cite{Kantawala02}, the authors propose another solution with a per-flow state and deficit round robin (DRR) scheduling to provide fairness guarantee. 
The main drawback of \cite{rai05}\cite{Kantawala02} is the need of a per-flow state while \cite{avrachenkov04} requires TCP senders modifications.
In \cite{pan00}, an active queue management algorithm, called CHOKe (CHOose and Keep for responsive flows), aims to approximate max-min fairness for the flows that cross a congested router. CHOKe is a stateless, simple to implement AQM and efficient scheme to identify and reduce the allocation of the flows which consume the most resources. 
The principle of this AQM is that, when a packet arrives at a congested router, CHOKe randomly chooses a packet from the FIFO buffer and compares it with an arriving packet. If they both belong to the same flow, they are both dropped. Although the goal of CHOKe is firstly to fairly share the capacity between flows, it might be considered as a solution for the problem of short flows from the point of view of the fairness.

\subsubsection{The case of giving a priority to certain TCP packets}

Giving a priority to certain TCP packets is not a novel idea. Several studies have tackled the benefit of this concept to improve the performance of TCP connection. This approach was really popular during the QoS networks research epoch as many queueing disciplines were enabled over IntServ and DiffServ testbeds allowing researchers to investigate such priority effects. Basically, the priority can be set intra-flow or inter-flow. 
Marco Mellia et al.~\cite{mellia02} have proposed to use intra-flow priority in order to protect some key identified packets from being lost of a TCP connection in order to increase the TCP throughput of a flow over an AF DiffServ class. In this study, the authors observe that TCP performance suffers significantly in the presence of bursty,  non-adaptive cross-traffic or when it operates in the small window regime, \textit{i.e.}, when the congestion window is small. The main argument is that bursty losses,  or losses during the small window regime, may cause retransmission timeouts (RTOs) which will result in TCP entering the slow-start phase. As a possible solution, the authors propose qualitative enhancements to protect against loss: the first several packets of the flow in order to allow TCP to safely exit the initial small window regime; several packets after an RTO occurs to make sure that the retransmitted packet is delivered with high probability and that TCP sender exits the small window regime; several packets after receiving three duplicate acknowledgement packets in order to protect the retransmission. This allows to protect the packets that strongly impact on the average TCP throughput against losses.
In \cite{chen03}\cite{Guo01}, the authors propose a solution on inter-flow priority. The short TCP flow are marked IN. Thus, packets from these flows are marked as a low drop priority. The differentiation in core routers is applied by an active queue management. When the sender has sent a number of packets that exceeds the flow identification threshold, the packet are marked OUT and the drop probability increases. However, these approaches need the support of a DiffServ architecture to perform \cite{rfc2475}.

\subsection{Acting at the TCP level}

The last solution is to act at the TCP level. The first possibility is to improve the behavior of TCP when a packet is dropped during this start up phase (\textit{i.e.} initial window size, limited transit). The second one is to prevent this drop by decreasing the probability of lost segments. For instance, in \cite{rfc5562}, the authors propose to apply an ECN mark to SYN/ACK segments in order to avoid to drop them. The main drawback of these solutions is that they require important TCP sender modifications that might involve heavy standardisation process. Some of these schemes can be seen as a complement of FavorQueue. For instance, we discuss the case of the increase of the initial slow-start value in Section \ref{sec:discuss}.

\section{FavorQueue description}
\label{sec:description}

Short TCP flows usually carry short TCP requests such as HTTP requests or interactive SSH or Telnet commands. As a result, their delay performance is mainly driven by:
\begin{enumerate}
  \item the end-to-end transfer delay. This delay can be reduced if the queueing delay of each router is low;
  \item the potential losses at the beginning connection. The first packets lost at the beginning of a TCP connection (\textit{i.e.} in the slow-start phase) are mainly recovered by the RTO mechanism. Furthermore, as the RTO is initially set to a high value, this greatly decreases the performance of short TCP flows. 
\end{enumerate}

The two main factors on which we can act to minimize the end to end delay and protect from loss the first packets of a TCP connection and are respectively the queuing delay and the drop ratio. Consequently, the idea we develop with FavorQueue is to favor certain packets in order to accelerate the transfer delay by giving a preferential access to transmission and to protect them from drop. 

This corresponds to implement a preferential access to transmission when a packet is enqueued and must be favored (temporal priority) and a drop protection is provided when the queue is full (drop precedence) with a push-out scheme that dequeues a standard packet in order to enqueue a favored packet.

When a packet is enqueued, a check is done on the whole queue to seek another packet from the same flow. If no other packet is found, it becomes a favored packet. The rationale is to decrease the loss of a retransmitted packet in order to decrease the RTO recovery ratio.
The proposed algorithm (given in Algorithm \ref{algo}) extends the one presented in \cite{dedu09pdp} by adding a drop precedence to non-favored packets in order to decrease the loss ratio of favored packets.
The selection of a favored packet is done on a per-flow basis. As a result the complexity is as a function of the size of the queue which corresponds to the maximum number of states that the router must handle. In order to select a packet, FavorQueue maintains an ordered linked list as a function of the number of packets belonging to a given flow. Knowing that 1)~a binary search is $log2(n)$ in terms of complexity; 2)~the insertion of a new element in a linked list is also of $log2(n)$ and 3) $n$ must be kept low \cite{Appenzeller04}, we can conclude that the process to select a packet is scalable\footnote{Another possible solution is to use a hash table which is of complexity O(1). However, the size of $n$ does not justify the implementation of such complex data structure.}.
    However the selection decision is local and temporary as the state only exists when at least one packet is enqueued. This explains why we prefer the term of ``favoring'' packets more than ``prioritizing'' them. Furthermore, FavorQueue does not introduced packet re-ordering inside a flow, which would obviously badly impacts TCP performance \cite{Arthur07}. 
Finally, in the specific case where all the traffic becomes favored, the behavior of FavorQueue will be identical to DropTail.

\begin{algorithm}[t]
\caption{FavorQueue algorithm}
\label{algo}
\begin{algorithmic}[1]
\STATE function enqueue(p)
\STATE \textit{\# A new packet p of flow F is received}
\IF {less than \textbf{1} packet of \textit{F} are present in the queue}
\STATE \textit{\# p is a favored packet}
	\IF {the queue is full}
		\IF {only favored packets in the queue}
 			\STATE p is drop
			\STATE return
		\ENDIF
        \ELSE
            	\STATE \textit{\# Push out}
		\STATE the last standard packet is dropped
	\ENDIF
	\STATE  p inserted before any standard packet\ 
\ELSE
	\STATE \textit{\# p is a standard packet}
	\IF {the queue is not full }
		\STATE p is put at the end of the queue
	\ELSE
		\STATE p is dropped
	\ENDIF 
\ENDIF 
\end{algorithmic}
\end{algorithm}

\section{Experimental methodology}
\label{sec:exp}

We use ns-2 to evaluate the performance of FavorQueue. Our simulation model allows to apply different levels of load to efficiently compare FavorQueue with DropTail. The evaluations are done over a simple dumbbell topology. The network traffic is modeled in terms of flows where each flow corresponds to a TCP file transfer. We consider an isolated bottleneck link of capacity C in bit per second. The traffic demand, expressed as a bit rate, is the product of the flow arrival rate $\lambda$ and the average flow size $E[\sigma]$. The load offered to the link is then defined by the following ratio:

\begin{equation}
\label{eq:rho}
\rho = \frac{\lambda E[\sigma]}{C}\textrm{.}
\end{equation}

The load is changed by varying the flow arrival rate \cite{Lachlan2008}. Thus, the congestion level increases as a function of the load. As all flows are independent, the flow arrivals are modeled by a Poisson process. 
A reasonable fit to the heavy-tail distribution of the flow size observed in practice is provided by the Pareto distribution. The shape parameter is set to $1.3$ and the mean size to 30 packets. 
Left side in Figure \ref{fig:delaypdf} gives the flows' size distribution used in the simulation model.

At the TCP flow level, the ns-2 TCP connection establishment phase is enabled and the initial congestion window size is set to two packets. As a result, the TCP SYN packet is taken into account in all dataset.
The load introduced in the network consists in several flows with different RTT according to the recommendation given in the "Common TCP evaluation suite" paper \cite{Lachlan2008}. The load is ranging from $0.05$ to $0.95$ in steps of $0.1$. The simulation is bounded to $500$ seconds for each given load. To remove both TCP feedback synchronization and phase effect \cite{Lachlan2008}, a traffic load of $10\%$ is generated in the opposite direction \cite{Lachlan2008}. The flows in the transient phase are removed from the analysis. More precisely, only flows starting after the first fifty seconds are used in the analysis. 
The bottleneck link capacity is set either to $10$\,Mb/s or $100$\,Mb/s. All other links have a capacity of $100$\,Mb/s (resp. $1000$\,Mb/s. According to the small buffers rule \cite{Ganjali06}, buffers can be reduced by a factor of ten. The rule-of-thumb says the buffer size $B$ can be set to $T \times C$ with $T$ the round-trip propagation delay and $C$ the link capacity. We choose $T=100$\,ms as it corresponds to the averaged RTT of the flows in the experiment. The buffer size at the two routers is set to a bandwidth-delay product with a delay of $10$\,ms. The packet length is fixed to $1500$ bytes and the buffer size has a length of 8 (resp. 83) packets. 

To improve the confidence of these statistical results, each experiment for a given load is done ten times using different sequences of pseudo-random numbers (in the following we talk about \textit{ten replications experiment}). Some figures also average the ten replications, meaning that we aggregate and average all flows from all ten replications and for all load conditions.
In this case, we talk about \textit{ten averaged experiment} results which represents a dataset of nearly 17 million of packets. The rationale is to consider these data as a real measurement capture where the load is varying as a function of time (as in \cite{ciullo09}) since each load condition has the same duration. In other words, this represents a global network behavior. 

The purpose of these experiments is to weight up the benefits brought by our scheme in the context of TCP best-effort flows. To do this, we first experiment a given scenario with DropTail then, we compare with the results obtained with FavorQueue. We enable FavorQueue only on the uplink (data path) while DropTail always remains on the downlink (ACK path).
We only compare all identical terminating flows for both experiments (\textit{i.e.} DropTail and FavorQueue) in order to assess the performance obtained in terms of service for a same set of flows.

We assume our model follows Internet short TCP flows characteristics as we found the same general distribution latency form as Figure \ref{fig:ciullo09} which is as a function of the measurements obtained in Figure \ref{fig:delaypdf}. This comparison provides a correct validation model in terms of latency. As explained above, Figure \ref{fig:delaypdf} corresponds and illustrates a \textit{ten averaged experiment}. 

To conclude, we recall in Table \ref{tab:exp}, the parameters used for each experiment. Note that for the experiment presented in Section \ref{subsec:persistent}, the duration is set to $2000$\,sec and there is no replication.

\begin{table*}[htb!]
	\begin{center}
	\newcolumntype{M}[1]{>{\centering}m{#1}}
	\setlength{\tabcolsep}{0.1cm} 
	\begin{tabular}{|c|c|M{7cm}|} 
	
		\cline{2-3} \multicolumn{1}{c|}{} & Parameters & Value \tabularnewline
		\hline \multirow{2}{*}{Simulation}  & Duration & $500$\,sec  \tabularnewline
		\cline{2-3}   & Replications & 10 \tabularnewline		
		\hline	
		\hline   & Queue Type & DropTail, RuN2C, LAS, Favor, CHOKe \tabularnewline
		\cline{2-3}         & Queue Size & $8$ packets ($83$ when C=$100$\,Mb/s)\tabularnewline
		\cline{2-3}   & RTT & from $4ms$ to $200$\,ms \tabularnewline
		\cline{2-3}   & Connecting links &  $100$\,Mb/s ($1000$\,Mb/s when C=$100$\,Mb/s)\tabularnewline		
		\cline{2-3}  \multirow{-5}{*}{Topology} & Core-link & $10$ or $100$\,Mb/s   \tabularnewline
		\hline 
		\hline   & TCP characteristics & TCP Newreno, $1500B$ of packet size, initial window$=2$ \tabularnewline		
		\cline{2-3}   & Flow length characteristics & Pareto, Shape$=1.3$, Average$=45000$\,B  \tabularnewline
		\cline{2-3}   & Load & from $0.05$ to $0.95$ (by step of $0.1$) \tabularnewline
		\cline{2-3}  \multirow{-5}{*}{Traffic} & Backward traffic & Load of $0.1$ \tabularnewline
		\hline
	\end{tabular}
	\caption{Summary of the experiments parameters}
	\end{center}
	\label{tab:exp}
\end{table*}

\section{Performance evaluation of TCP flows with FavorQueue}
\label{sec:perf}

We present in this section the global performance obtained by FavorQueue. We then analyze its performance deeper and investigate the case of persistent flows. We compare a same set of flows to assess the performance obtained with DropTail and FavorQueue.

\subsection{Overall performance}
\label{subsec:overall}

We are interested in assessing the performance of each TCP flows in terms of latency, drop ratio and goodput.
We recall from Section \ref{sec:intro} that we defined the latency as the time to complete a data download (\textit{i.e.} the transmission time) and the goodput is the average pace of the download.
In order to assess the overall performance of FavorQueue compared to DropTail and other AQMs that target a similar goal as our proposal, 
we evaluate FavorQueue against other mechanisms introduced in Section \ref{sec:related} and in particular: LAS scheduling mechanism \cite{Kleinrock75}, RuN2C scheduling policy \cite{avrachenkov04} and finally against CHOKe \cite{pan00}.

\begin{figure}[htb!]
	\centering
	\subfigure[Bottleneck set to 10Mb/s]{
                \includegraphics[width=0.45\textwidth]{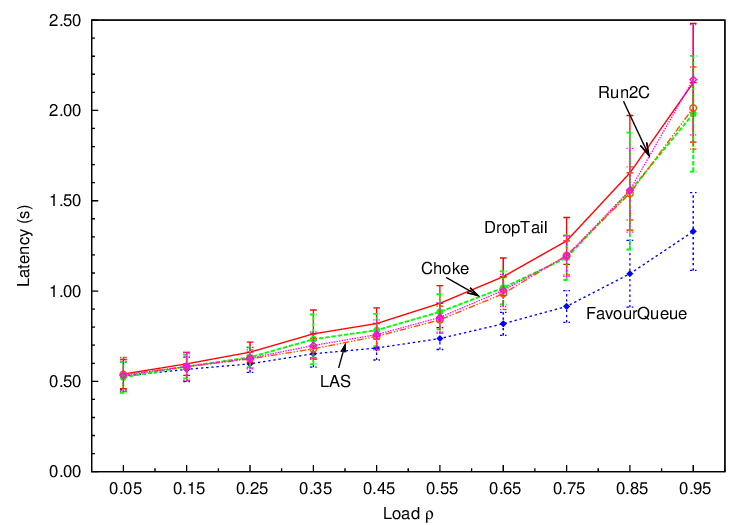}
                \label{fig:load10}
		}
	\subfigure[Bottleneck set to 100Mb/s]{                
		\includegraphics[width=0.45\textwidth]{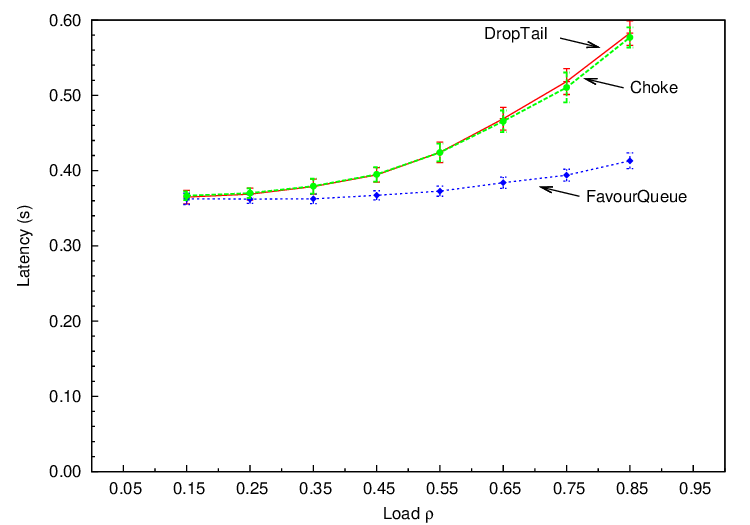}
                \label{fig:load100}
		}
	\caption{Overall latency according to traffic load.}
	\label{fig:load}
\end{figure}

\begin{figure}[htb!]
	\centering
	\subfigure[Bottleneck set to 10Mb/s]{
                \includegraphics[width=0.45\textwidth]{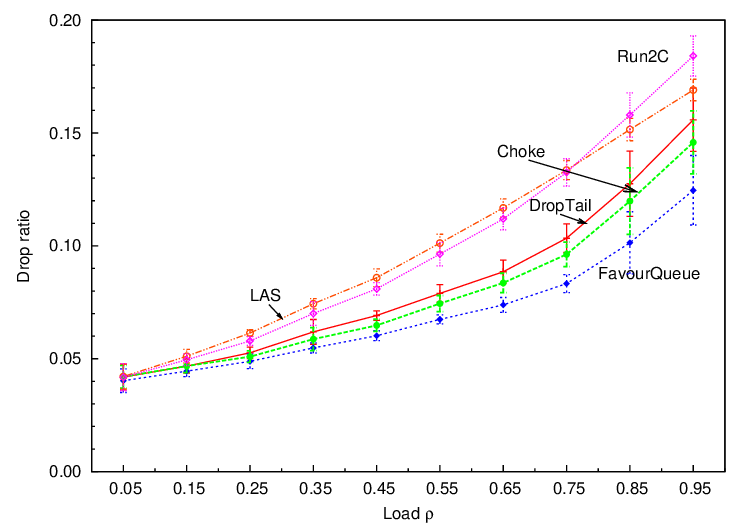}
                \label{fig:drop10}
		}
	\subfigure[Bottleneck set to 100Mb/s]{
                \includegraphics[width=0.45\textwidth]{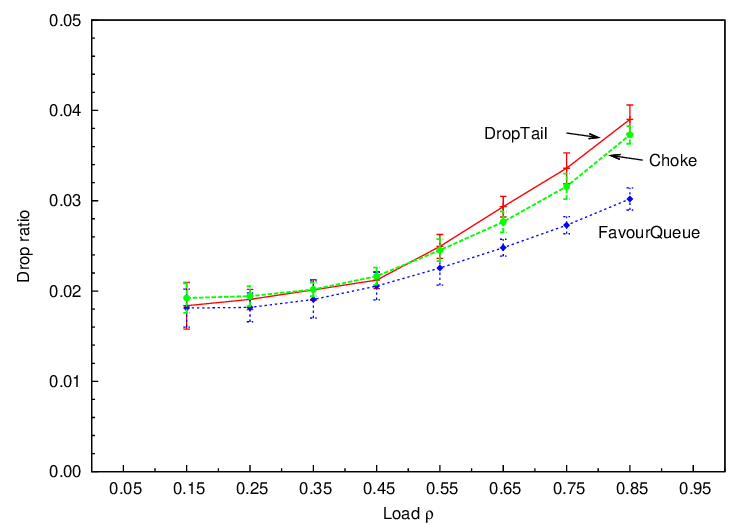}
                \label{fig:drop100}
		}
	\caption{Overall drop ratio according to traffic load.}
	\label{fig:drop}
\end{figure}

The parameters used for these AQMs are those proposed by their authors. In particular for RuN2C, the authors claim that in the context of TCP, the threshold parameter can be set to a small value (we choose to set this threshold to twenty packets) in order to guarantee that a TCP connection will recover from a loss by fast retransmit rather than a timeout.
The results obtained are presented in Figure \ref{fig:load}, concerning the mean and standard deviation of the latency as a function of the traffic load of FavorQueue and in Figure \ref{fig:drop}, concerning the drop ratio.
These results are unequivocal. FavorQueue provides a gain when the load increases compared to all other AQM (\textit{i.e.} when the queue has a significant probability of having a non-zero length) while the drop precedence clearly brings out a significant gain in terms of latency.

\begin{figure}[htb!]
	\centering
	\subfigure[DropTail]{
                \includegraphics[width=0.30\textwidth]{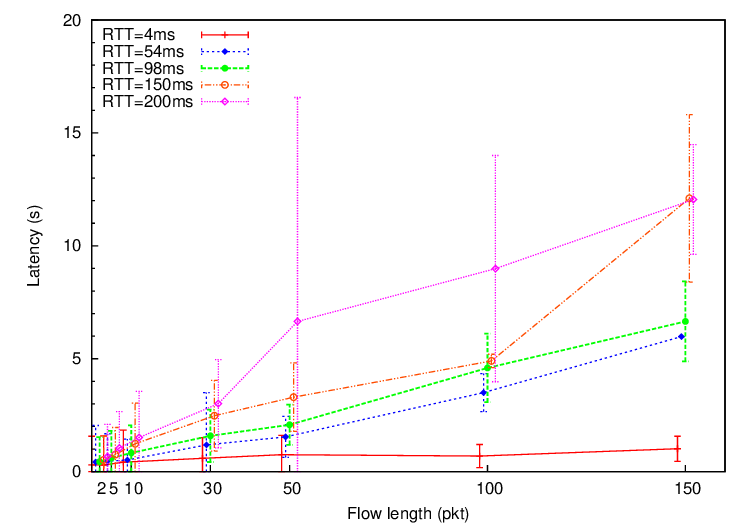}
                \label{fig:rttDT}
		}
	\subfigure[CHOKe]{                
		\includegraphics[width=0.30\textwidth]{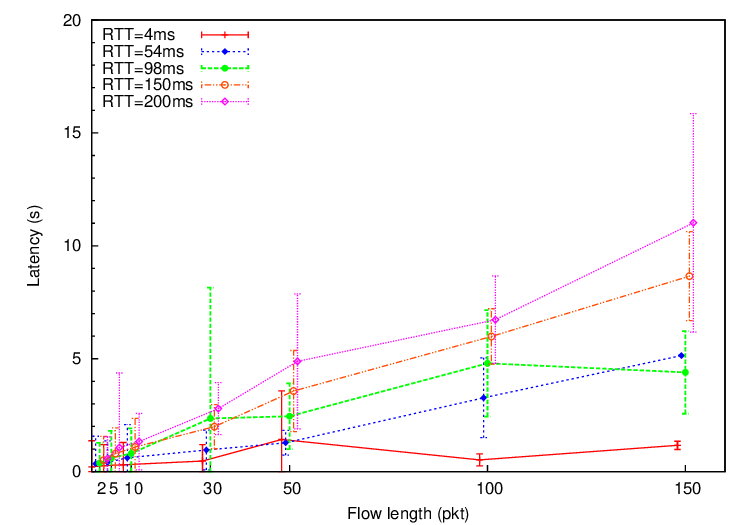}
                \label{fig:rttCK}
		}
	\subfigure[FaQ]{                
		\includegraphics[width=0.30\textwidth]{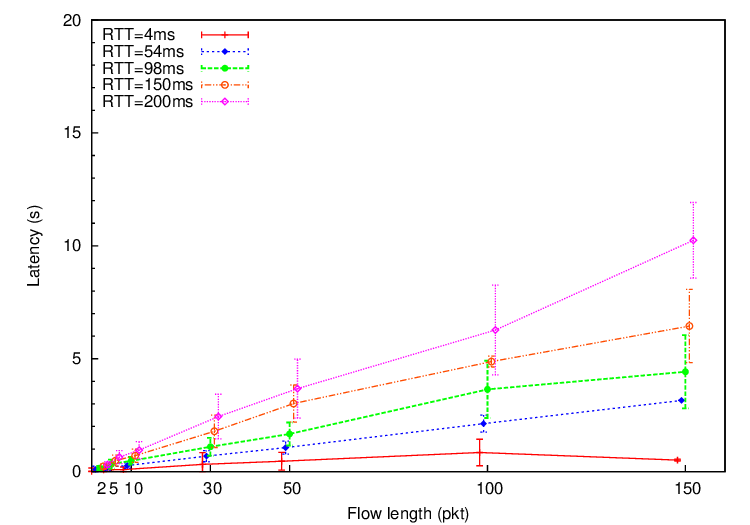}
                \label{fig:rttFAQ}
		}
	\caption{Latency according to flow length and various RTT.}
	\label{fig:RTTQ}
\end{figure}

Figures \ref{fig:load10} and \ref{fig:drop10} show that RUN2C loses its efficiency in heavy load condition. In this case, there are several short flows and the action of RUN2C becomes too deterministic. In comparison, CHOKe that acts randomly, obtained a better drop ratio. 
In fact, RUN2C tends to favor all the early first TCP segments (i.e. in our simulation, all the first fifteen ones). When there are several short flows, long ones are never favored. This is not the case with CHOKe which enables a probabilist action.
FavorQueue does not perform as RUN2C and still discriminates flows in heavy load condition while preventing the loss of a retransmitted packet. This explains the delay improvement observed in Figure \ref{fig:load10} compared to CHOKe.
For the sake of completion, we also provide an experiment with a bottleneck capacity of $100$\,Mb/s in Figures \ref{fig:load100} and \ref{fig:drop100}. In order to ease the reading, we only provide the results for DropTail, CHOKe and FavorQueue. We observe that these results are homothetic with those obtained with the $10$\,Mb/s bottleneck capacity. In the next experiments of this section, we choose to report only the results of these three AQMs to ease the reading.

Following this, we have computed the resulting normalized goodput for all flows size for all experiments and obtained is 2.4\% with DropTail and 3.5\% with FavorQueue (\textit{i.e.} around 1\% of difference). This value is not weak as it corresponds to an increase of 45\%.

Figure \ref{fig:RTTQ} complete these measurements by giving the latency obtained by DropTail, CHOKe and FavourQueue according to flow length and various RTT. These curves highlight another good property of the proposed AQM. Compare to CHOKe and DropTail, FavourQueue provide smoother results. 

\begin{figure}[htb!]
   \begin{minipage}[b]{1.0\columnwidth}
	\centering
	\includegraphics[width=0.7\columnwidth]{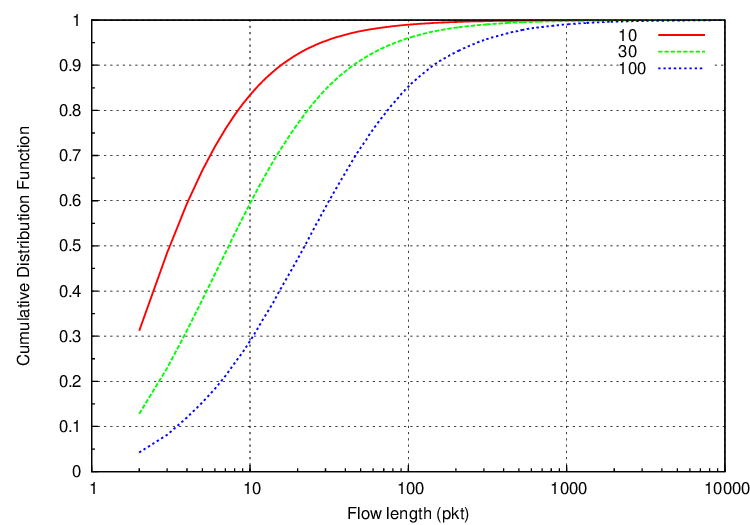}
	\caption{Traffic Distribution.}
	\label{fig:size-dist}
   \end{minipage}\hfill

   \begin{minipage}[b]{1.0\columnwidth}   
	\centering
	\includegraphics[width=0.7\columnwidth]{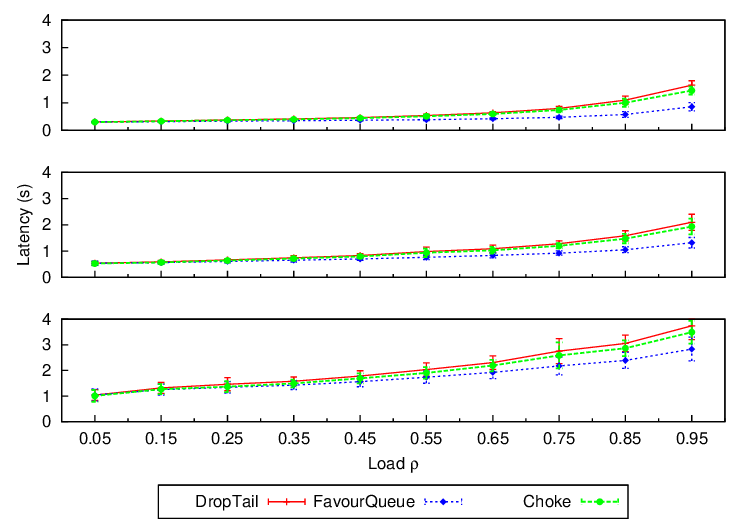}
	\caption{Latency for each traffic distribution.}
	\label{fig:lat-dist}
   \end{minipage}
   \label{fig:dist}
\end{figure}

We propose to extend these results with three experiments corresponding to three Pareto traffic distribution centered on 10, 30 and 100 represented in Figure \ref{fig:size-dist}. The objective is to assess the impact of the Pareto distribution on the latency results. These three values can simulate various traffic ranging from home access to datacenters for instance. The results obtained are given only for CHOKe, FaQ and DT in Figure \ref{fig:lat-dist} to ease the reading. As shown in this figure, we obtained homothetic results with those previously presented.

\begin{figure}[htb!]
   \begin{minipage}[b]{1.0\columnwidth}   
	\centering
	\includegraphics[width=0.7\columnwidth]{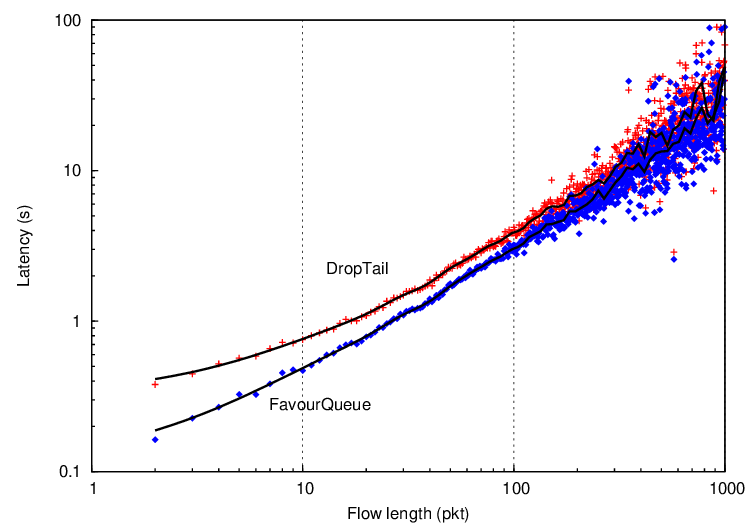}
	\caption{Mean latency as a function of the flow length.}
	\label{fig:latencysize}
   \end{minipage}

   \begin{minipage}[b]{1.0\columnwidth}
	\centering
	\includegraphics[width=0.7\columnwidth]{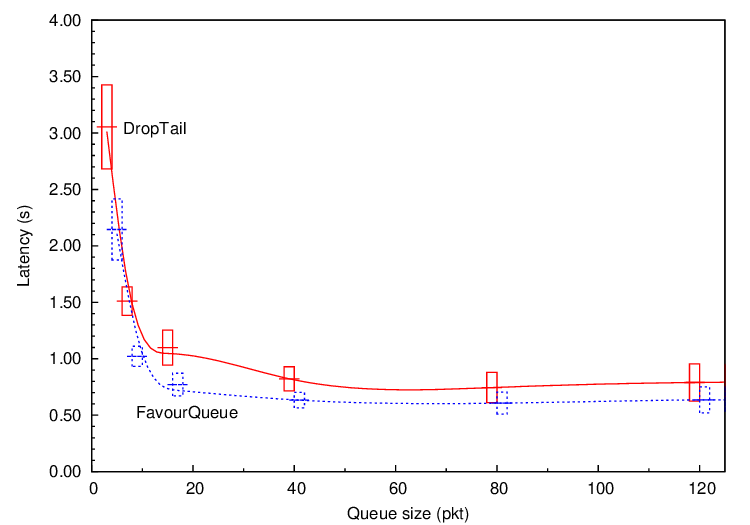}
	\caption{Overall latency according to queue size.}
	\label{fig:latencyqueue}
   \end{minipage}\hfill
\end{figure}

Figure \ref{fig:latencysize} gives the average latency as a function of the flow length.
On average, we observe that FavorQueue obtains a lower latency than DropTail whatever the flow length. This difference is also larger for the short TCP flows which are also numerous (we recall that the distribution of the flows' size follows a Pareto distribution and as a result the number of short TCP flow is higher). This demonstrates that FavorQueue particularly favors the slow-start of every flow and as a matter of fact, short TCP flows. The cloud pattern obtained for a flow size higher than a hundred is due to the decrease of the statistical sample (following the Pareto distribution used for the experiment) that result in a greater dispersion of the results obtained.
As a result, we cannot drive a consistent latency analysis for sizes higher than hundred.

To complete these results, Figure \ref{fig:latencyqueue} gives the latency obtained when we increase the queue size. We observe that whatever the queue size, FavorQueue always obtains a lower latency. Beyond a given queue size (in Figure \ref{fig:latencyqueue} at $x=60$), the increase of the queue does not have an impact on the latency. 
This enforces the consistency of the solution as Internet routers prevent the use of large queue size \cite{Appenzeller04}. In the following, the queue size is set to $8$ packets.

\subsection{Performance analysis}

\begin{figure}[ht]
\begin{center}
\includegraphics[width=0.7\columnwidth]{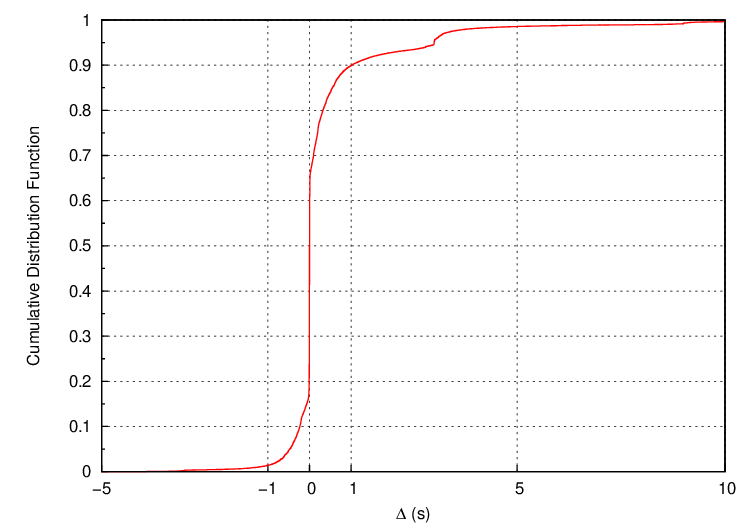}
\end{center}
\caption{Cumulative distribution function of latency difference $\Delta$.}
\label{fig:diffcdf}
\end{figure}

To refine our analysis of the latency, we propose to evaluate the difference of latencies per flows for both queues. We denote $\Delta{i} = Td_{i}-Tf_{i}$ with $Td$ and $Tf$ the latency observed respectively by DropTail and FavorQueue for a given flow~$i$. Figure \ref{fig:diffcdf} gives the cumulative distribution of the latencies difference.
This figure illustrates that there is more decrease of the latency for each flow than increase. Furthermore for 16\% of flows, there is no impact on the latency \textit{i.e.} $\Delta = 0$. In other words, 84\% of flows observe 
a change of latency; 54\% of flows observe a decrease ($\Delta > 0$) and 10\% of flows observe a significant change ($\Delta > 1$ second). However, 30\% of the flows observe an increase of their latency ($\Delta < 0$).
Note that the variation below 10ms are not visible in this figure. As 18\% of the flows have $\Delta{i} < -10ms$ and 64\% have $\Delta{i} < 10ms$, the percentage of flows between  18\% and 64\% represent 46\%.
In summary, FavorQueue has a positive impact on certain flows that are penalised with DropTail.

\begin{figure}[ht]
\begin{center}
\includegraphics[width=0.7\columnwidth]{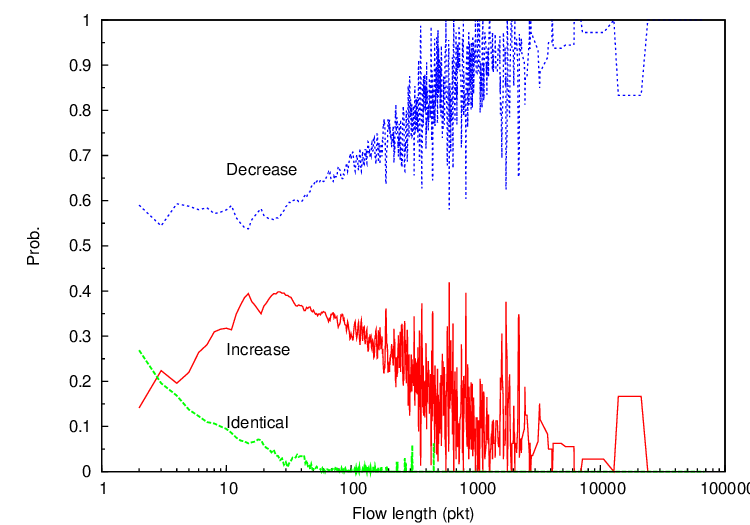}
\end{center}
\caption{Probability to change the latency.}
\label{fig:diffprob}
\end{figure}

In order to assess the flows that obtain a lower latency, Figure \ref{fig:diffprob} gives the probability of latency improvement. 
For the whole set of short TCP flows, (\textit{i.e.} with a size lower than 10 packets), the probability to improve the latency reaches 58\% while the probability to decrease is 25\%. For long TCP flows (\textit{i.e.} above 100 packets), 
the probability to improve and to decrease the latency is respectively 80\% and 20\%.
The flows with a size around 30 packets are the ones with the highest probability to be penalised. 
For long TCP flows, the large variation of the probability indicates a uncertainty which mainly depends on the experimental conditions of the flows.
We have to remark that long TCP flows are less present in this experimental model (approximately 2\% of the flows have a size higher or equal to 100 packets). As this curve corresponds to a ten averaged experiment, 
each long TCP flows have experienced various load conditions and this explains these large oscillations. 

Medium sized flows are characterized by a predominance of the slow-start phase. During this phase, each flow opportunistically occupies the queue and as a results less packets are favored due to the growth of the TCP window. The increase of the latency observed for medium sized flows (ranging from 10 to 100) is investigated later in subsection \ref{subsec:medium}. We will also see in the next subsection \ref{subsec:persistent} that FavorQueue acts like a shaper for these particular flows.

\begin{figure}[ht]
\begin{center}
\includegraphics[width=0.7\columnwidth]{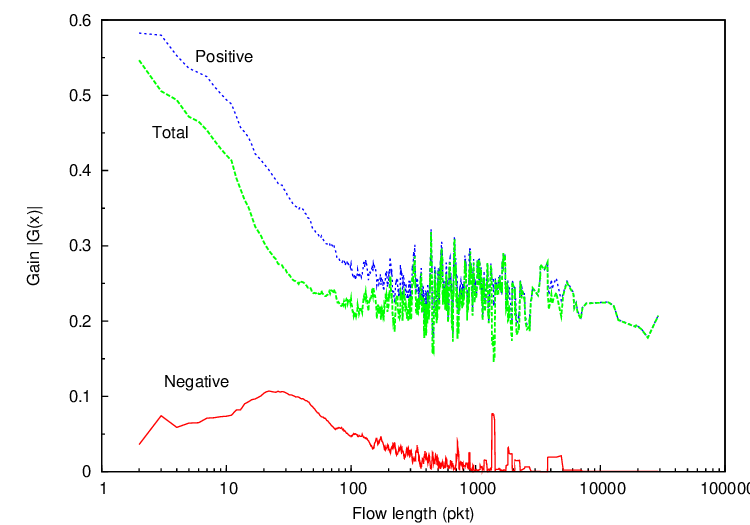}
\end{center}
\caption{Average Latency gain per flow length.}
\label{fig:diffgain}
\end{figure}

To estimate the latency variation, we define $G(x)$ the latency gain for the flows of length $x$ as follows:

\begin{equation}
G(x) = \frac{\sum_{i=1}^{N} \Delta_{x_{i}} }{\sum_{i=1}^{N}Td_{x_{i}}}\textrm{.}
\end{equation}

with $N$, the number of flows of length $x$. A positive gain indicates a decrease of the latency with FavorQueue. 
Figure \ref{fig:diffgain} provides the positive, negative and total gains as a function of the flows size. We observe an important total gain for the short TCP flows. 
The flows with an average size obtain the highest negative gain and this gain also decreases when the size of the flows increases.
Although some short flows observe an increase of their latency, in a general manner, the positive gain is always higher. 
This preliminary analysis illustrates that FavorQueue improves by 30\% on average the best-effort service in terms of latency. The flows that take the biggest advantage of this scheme are the short flows with a gain up to 55\%.

\begin{figure}[ht]
\begin{center}
\includegraphics[width=0.7\columnwidth]{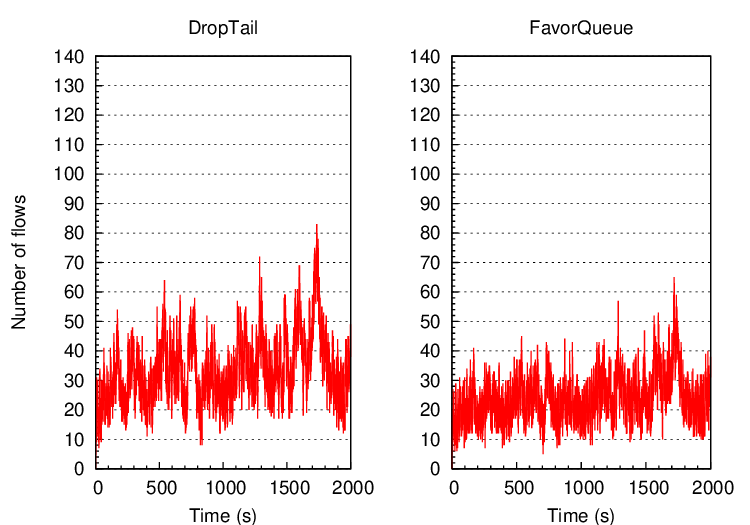}
\end{center}
\caption{Number of simultaneous flows in the network.}
\label{fig:session}
\end{figure}

Finally and to conclude with this section, we plot in Figure \ref{fig:session} the number of flows in the system under both AQM as a function of time 
to assess the change in the stability of the network. 
We observe that FavorQueue considerably reduces both the average number of flows in the network as well as the variability. 

\subsection{The case of persistent flows}
\label{subsec:persistent}

\begin{figure}[htb!]
\begin{center}
\includegraphics[width=0.7\columnwidth]{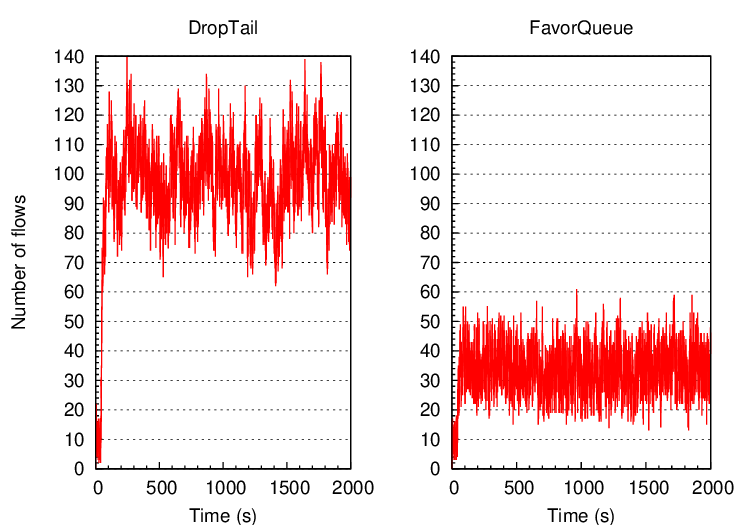}
\end{center}
\caption{Number of short flows in the network when persistent flows are actives.}
\label{fig:sessionl}
\end{figure}

\begin{figure}[htb!]
   \begin{minipage}[b]{1.0\columnwidth}
	\centering
	\includegraphics[width=0.7\columnwidth]{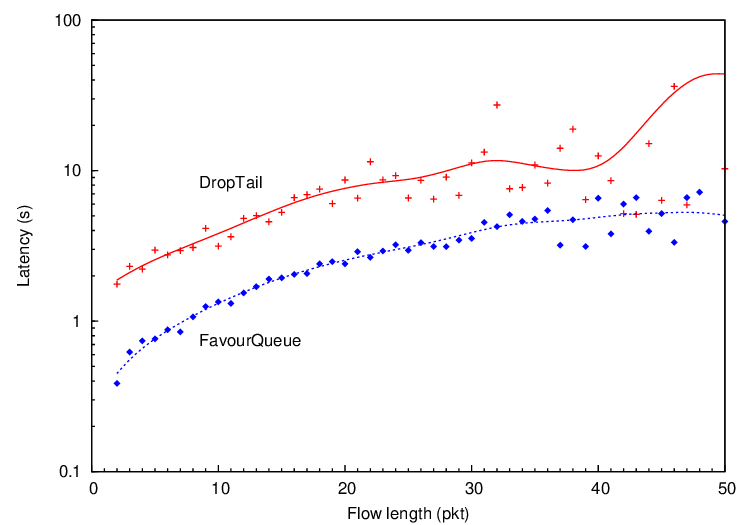}
	\caption{Mean latency as a function of flow size for short flows in presence of persistent flows.}
	\label{fig:platency}
   \end{minipage}

   \begin{minipage}[b]{1.0\columnwidth}   
	\centering
	\includegraphics[width=0.7\columnwidth]{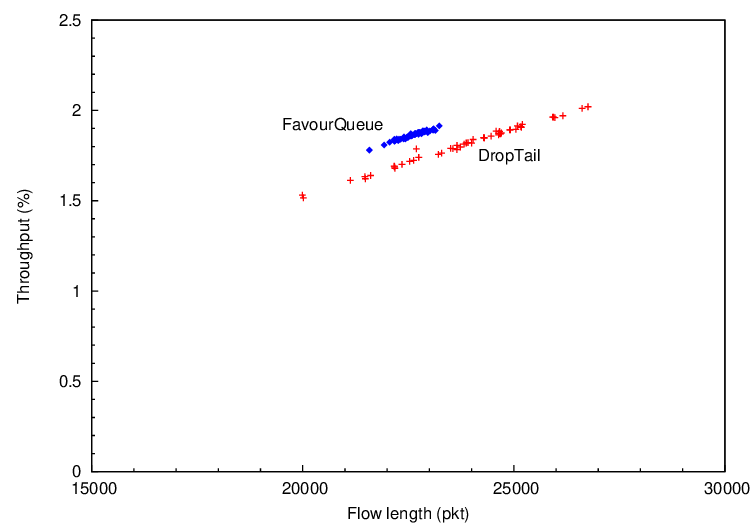}
	\caption{Mean throughput as a function of flow length for 50 persistent flows.}
	\label{fig:pthru}
   \end{minipage}
\end{figure}

Following \cite{avrachenkov04} whence we borrow the same experimental scenario and hypothesis, we evaluate how the proposed scheme affects persistent flows with randomly arriving short TCP flows.
We now change the network conditions with 20\% of short TCP flows with exponentially distributed flow sizes with a mean of 6 packets. Forty seconds later, 50 persistent flows are sent. Figure \ref{fig:sessionl} gives 
the number of simultaneous short flows in the network. When the 50 persistent flows start, the number of short flows increases and oscillates around 100 with DropTail. By using FavorQueue, the number increases to 30 short flows. 
The short flows still take advantage of the favor scheme and Figure \ref{fig:platency} confirms this point. However we observe in Figure \ref{fig:pthru} that the persistent flows are not penalized. The mean throughput is 
nearly the same (1.81\% for DropTail versus 1.86\% for FavorQueue) and the variance is smaller with FavorQueue. 
Basically, FavorQueue acts as a shaper by slowing down opportunistic flows while decreasing the drop ratio of non opportunistic flows (those which less occupy the queue). 

\section{Understanding FavorQueue}
\label{sec:understand}

The previous section has shown the benefits obtained with FavorQueue in terms of service. 
In this section, we analyse the reasons of the improvements brought by FavorQueue by looking at the AQM performance. We study the drop ratio and the queueing delay obtained by both queues in order to assess the reasons of the gain obtained by FavorQueue. We recall that for all experiments, FavorQueue is only set on the upstream. The reverse path uses a DropTail queue. In a first part, we look at the impact of the AQM on the network then on the end-host.

\subsection{Impact on the network}
\label{subsec:medium}

\begin{figure}[htb!]
   \begin{minipage}[b]{1.0\columnwidth}
	\centering
	\includegraphics[width=0.7\columnwidth]{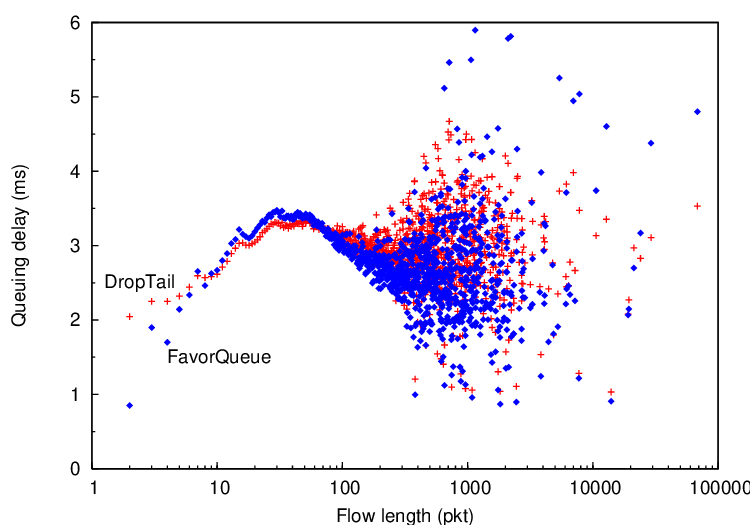}
	\caption{Average queuing delay according to flow length.}
	\label{fig:queuedelay}
   \end{minipage}

   \begin{minipage}[b]{1.0\columnwidth}   
	\centering
	\includegraphics[width=0.7\columnwidth]{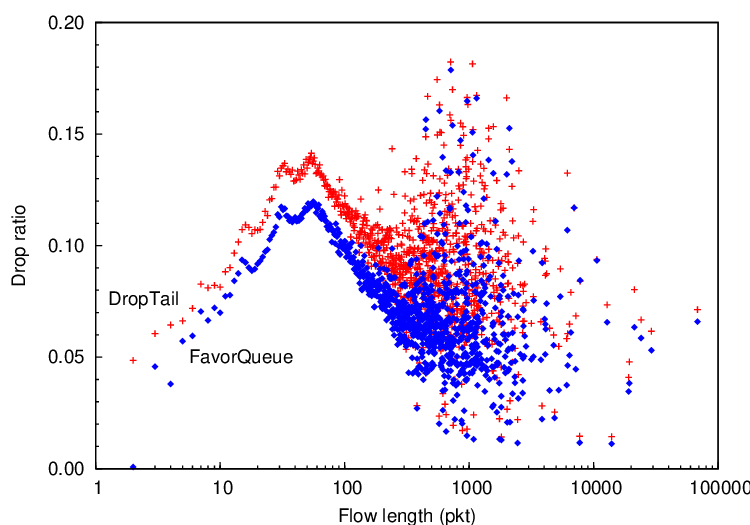}
	\caption{Average drop ratio according to flow length.}
	\label{fig:dropavg}
   \end{minipage}
\end{figure}

Figure \ref{fig:queuedelay} shows the evolution of the average queueing delay depending on the size of the flow. This figure corresponds to the 10 averaged replications experiment (as defined Section \ref{sec:exp}).
Basically, the results obtained by FavorQueue and DropTail are similar. Indeed, the average queueing delay is 2.8ms for FavorQueue versus 2.9ms for DropTail and both curves similarly behave. 
We can notice that the queueing delay for the medium sized flows slightly increases with FavorQueue. These flows are characterized by a predominance of the slow-start phase as most of the packets that belong to these flows are emitted during the slow-start. Since during this phase each flow opportunistically occupies the queue, less packets are favored due to the growth of the TCP window. As a result, their queuing delay increases. When the size of the flow increases (above a hundred packets length), the slow-start is not pervasive anymore and the average queueing delay of each packet of these flows tends to be either higher or lower as suggested by the cloud Figure \ref{fig:queuedelay} depending on the number of favored packets during their congestion avoidance phase.

However and as Figure \ref{fig:dropavg} suggests, the good performance in terms of latency obtained by FavorQueue (previously shown in Figure \ref{fig:load} from Section \ref{sec:perf}) is mostly due to a 
significant decrease of the drop ratio. If we look at the average drop ratio of both queues in Figure \ref{fig:dropavg}, still as a function of the flow length,  
we clearly observe that the number of packets dropped is lower for FavorQueue. Furthermore, the loss ratio for the flow size of 2 packets is about $10^{-3}$ meaning that the flows of this size obtain a benefit compared to DropTail.
The slow-start phase is known to send bursts of data \cite{Barakat00}. Thus, most of the packets sent during the slow-start phase have a high probability to be not favored. This explains the increase of the drop ratio according 
to the flow size until 60 packets.
Indeed, in the slow-start phase, packets are sent by burst of two packets. As a result, the first packet is favored and the second
one will be favored only if the first is already served. Otherwise, the second packet might be delayed. Then, when their respective acknowledgements are back to the source, the next sending will be more spaced. 
As FavorQueue might decrease the burstiness of the slow-start, we might decrease the packet loss rate and thus improve short TCP flow performance.

If we conjointly consider both figures \ref{fig:queuedelay} and \ref{fig:dropavg}, we observe that FavorQueue enables a kind of traffic shaping that decreases TCP aggressivity during the slow-start phase which results in a 
decrease of the number of dropped packets. As the TCP goodput is proportional to $1/(RTT.\sqrt{p})$ \cite{mathis97macroscopic}, the decrease of the drop ratio leads to an increase of the goodput which explains the good 
performance obtained by FavorQueue in terms of latency.

The loss ratio of SYN segments is on average 1.8\% with DropTail. However for a load higher than 0.75, this loss ratio value reaches 2.09\% while with FavorQueue, this ratio is 0.06\%. Finally on average for all load conditions, this value is 0.04\% with FavorQueue. These results demonstrate the positive effect of protecting SYN segments from being dropped. By using FavorQueue in duplex mode (we recall that we have tested FavorQueue only on the upstream), this would further improve the results as SYN/ACK packets would have also been protected.

\subsection{Impact on the end-host performance}

\begin{figure}[htb!]
   \begin{minipage}[b]{1.0\columnwidth}
	\centering
	\includegraphics[width=0.7\columnwidth]{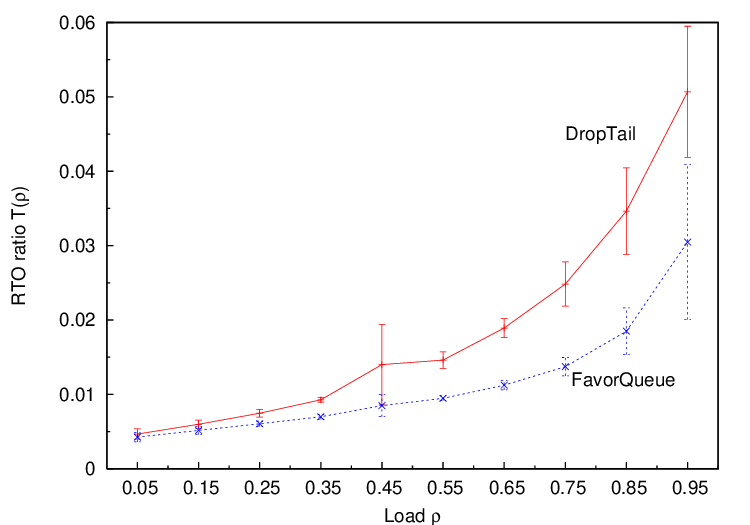}
	\caption{RTO ratio as a function of the network load.}
	\label{fig:rtorate}
   \end{minipage}

   \begin{minipage}[b]{1.0\columnwidth}   
	\centering
	\includegraphics[width=0.7\columnwidth]{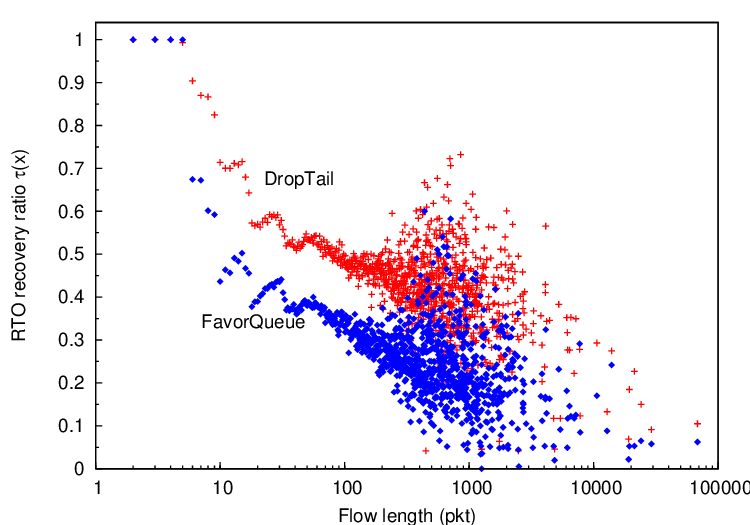}
	\caption{RTO recovery ratio according to flow length.}
	\label{fig:rtorec}
   \end{minipage}
\end{figure}

The good performance obtained with FavorQueue in terms of latency are linked to the decrease in losses at the start of the flow.
In the following, we propose to estimate the benefits of our scheme by estimating the RTO ratio as a function of the network load. 
We define the RTO ratio $T(\rho)$ for a given load $\rho$ as follows:
\begin{equation}
\label{eq:rto}
T(\rho) = \frac{\sum_{i=1}^{N} {RTO}_{i}} {\sum_{i=1}^{N} (L_{i}+R_{i})},
\end{equation}
with ${RTO}_{i}$ the number of RTO for the $i^{th}$ flow; $R_i$ its number of retransmissions and $L_{i}$ the size of flow $i$.
The ten replications experiment in Figure \ref{fig:rtorate} presents the evolution of the RTO ratio for FavorQueue and DropTail and shows that the decrease of the loss ratio results 
in a decrease of the RTO ratio for FavorQueue. This also shows the advantage to use FavorQueue when the network is heavily loaded.

We now evaluate the RTO recovery ratio as a function of the flow length. We define this RTO recovery ratio as follows:
\begin{equation}
\label{eq:taux}
\tau(x) = \frac{\sum_{i=1}^{N} {RTO}_{i}} {\sum_{i=1}^{N} ({RTO}_{i}+ {FR}_{i})},
\end{equation}
with ${FR}_{i}$ the number of TCP Fast Retransmits for the $i^{th}$ flow. In terms of RTO recovery, Figure \ref{fig:rtorec} shows a significant decrease of 
the number of recoveries with an RTO. Concerning the ratio of Fast Retransmits for this experiment, we observe an increase of 14\% with FavorQueue. 
As a fast recovery packet is placed at the beginning of a window, FavorQueue prevents the loss of a retransmission. 
Then, the number of recovery with Fast Retransmit is higher with FavorQueue and the latency observed is better since the retransmission are faster.

The objective of Figure \ref{fig:rtorec} is to quantify the number of RTO among the number of TCP recovery. Indeed, a Fast Retransmit recovery is always better in terms of delay than a timeout (RTO) recovery. Thus, this metric allows to assess how the recovery is sliced between FR and RTO. The first figure \ref{fig:rtorate} gives a quantitative information while the second one \ref{fig:rtorec} gives a qualitative information on the percentage of recovery between FR and RTO.

For flows with a size strictly below six packets, the recovery is exclusively done by RTO. Indeed, in this case there is not enough duplicate acknowledgements to trigger a Fast Retransmit.
For flows above six packets, we observe a noticeable decrease of the RTO ratio due to the decrease of the packet lost rate on the first packets of the flow. Thus, the number of duplicate acknowledgement is higher, allowing to
trigger a Fast Retransmit recovery phase. The trend shows a global decrease of the RTO ratio when the flow length increases. 
On the overall, the RTO recovery ratio reaches 56\% for DropTail and 38\% for FavorQueue.
The decrease of the gain obtained follows the increase of the flow size. This means that FavorQueue helps the connection establishment phase.

\section{Stochastic model of FavorQueue}
\label{sec:model}

We analyze, in this part, the impact of the temporal and drop priorities previously defined in Section \ref{sec:description}. We also propose a stochastic model of the mechanism to better understand some results presented. 

\subsection{Preliminary statistical analysis}

\begin{figure}[htb!]
   \begin{minipage}[b]{1.0\columnwidth}
	\centering
	\includegraphics[width=0.7\columnwidth]{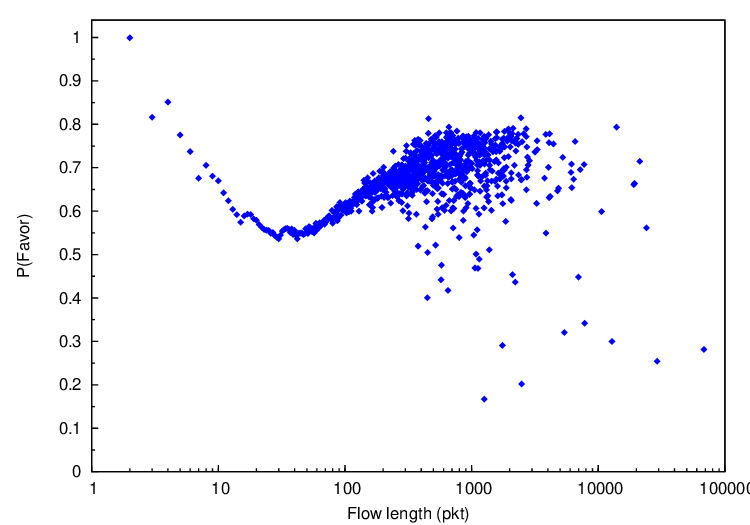}
	\caption{Probability of packet favoring according to flow length.}
	\label{fig:probaction}
   \end{minipage}

   \begin{minipage}[b]{1.0\columnwidth}   
	\centering
	\includegraphics[width=0.7\columnwidth]{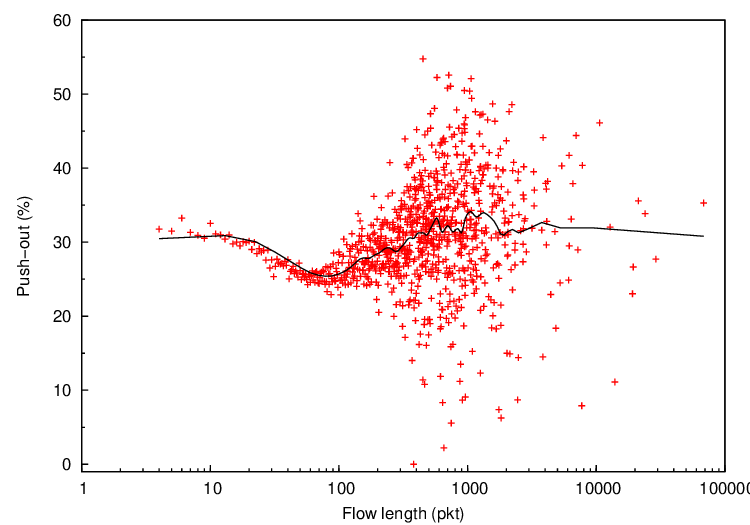}
	\caption{Push-out proportion of drop as a function of flow length.}
	\label{fig:propo}
   \end{minipage}
\end{figure}

We first estimate the probability of favoring a flow as a function of its length by a statistical analysis. We define $P(Favor|S=s)$, the probability to favor a flow of size $s$, as follows:

\begin{equation}
\label{eq:favor}
P(Favor|S=s) = \frac{\sum_{i=1}^{N} fa_{i} }{\sum_{i=1}^{N} s + R_{i}}\textrm{.}
\end{equation} 

with $fa_{i}$, the number of packets which have been favored and $R_{i}$ the number of retransmitted packets of a given $i$ flow.
The number of favored packets corresponds to the number of packets selected to be favored at the router queue.
Figure \ref{fig:probaction} gives the results obtained and shows that:

\begin{itemize}
\item flows with a size of two packets are always favored; 
\item middle sized flows that mainly remain in a slow-start phase are less favored compared to short flows. The ratio reaches 50\% meaning that one packet out of two is favored; 
\item long TCP flows get a favoring ratio around 70\%. 
\end{itemize} 

We also investigate the ratio of packets dropped resulting from the push-out algorithm as a function of the flow length in order to assess whether some flows are more penalised by push-out.
As shown, Figure \ref{fig:propo}, the mean is about 30\% for all flows, meaning that the push-out algorithm does not impact short than long TCP flows.

We now propose to build a stochastic model to explain and better understand the shape of Figure \ref{fig:probaction} in the following.

\subsection{Stochastic model}

We denote $S$: the random variable of the size of the flow and $Z$: the Bernoulli random variable which is equal to $0$ if no favored packets are present in the queue and $1$ otherwise.
We then distinguish three different phases:
\begin{itemize}
	\item phase \#1 : all flows have a size smaller than $s_{1}$. In this phase, the flows are in slow-start mode. This size is a parameter of the model which depends of the load. ;
	\item phase \#2 : all flows have a size larger than $s_{1}$ and smaller than $s_{2}$.  In this phase, flows progressively leave the slow-start mode (corresponding to the  bowl between $[10:100]$ in Figure \ref{fig:probaction}). This is the most complex phase to model as all flows are either in the congestion avoidance phase or at the end of their slow-start. $s_{2}$ is also a parameter of the model which depends of the load;
	\item phase \#3 : all flows have a size larger than $s_{2}$. All flows are in congestion avoidance phase. Note that the statistical sample which represents this cloud is not large enough to correctly model this part (as already pointed out in Section \ref{subsec:overall}). However, one other important result given by Figure \ref{fig:probaction} is that 70\% of the packets belonging to flows in congestion avoidance mode are favored. We will use this information to infer the model. This also confirms that the spacing between each packet in the congestion avoidance phase increases the probability of an arriving packet to be favored. 
\end{itemize}

\subsubsection*{First phase}

We consider a bursty arrival and assume that all packets belonging to the previous RTT have left the queue. Then, the burst size (BS) can take the following values: $BS=1, 2, 4, 8, 16, 32, ...$. If $Z=0$, we assume that a maximum of 3 packets can be favored in a row\footnote{The rationale is the following, if $Z=0$ a single packet (such as the SYN packet) is favored and one RTT later, the burst of two packets (or larger) will be favored if we consider that the first packet of this burst is directly served.}. The packets number that are favored in this case are $1, 2, 3, 4, 5, 6, 8, 9, 10, 16, 17, 18, ...$ and  $1, 2, 4, 8, 16, 32, ...$ if $Z=1$. Thus, if $Z=0$, the probability to favor a packet of a flow of size $s$ is:

\begin{equation}
\label{eq1}
P(Favor|(Z=0,S=s)) = \left \lbrace
\begin{array}{c}
1,s \leqslant 6 \\
\frac{s-1}{s}, 7 \leqslant s \leqslant 10 \\
\frac 9 s, 11 \leqslant s \leqslant 15 \\
\frac{s-6}{s}, 16 \leqslant s \leqslant 18 \\
\frac{12}{s}, 19 \leqslant s \leqslant 31 \\
(...)
\end{array} \right.
\end{equation}

and with $Z=1$:

\begin{equation}
\label{eq2}
P(Favor|(Z=1,S=s)) = \left \lbrace
\begin{array}{c}
1, s=1 \\
\frac 2 s, 2 \leqslant s \leqslant 3 \\
\frac 3 s, 4 \leqslant s \leqslant 7 \\
\frac 4 s, 8 \leqslant s \leqslant 15 \\
\frac 5 s, 16 \leqslant s \leqslant 31 \\
(...)
\end{array} \right.
\end{equation}

The probability to favor a packet of a flow of size $s$ is thus: 
\small
\begin{eqnarray}
\label{eqn_cond1}
P(Favor | S=s) = & P(Z=0).P(Favor|(Z=0,S=s))\\ \nonumber
& + P(Z=1).P(Favor|(Z=1,S=s))
\end{eqnarray}
\normalsize

Once again, $P(Z=0)$ and $P(Z=1)$ depend on the load of the experiment and must be given.

\subsubsection*{Second phase}

In this phase, each flow progressively leaves the slow-start phase. First, when a flow finishes its slow-start phase, each following packet has a probability to be favored of $70\%$ (as shown in in Figure \ref{fig:probaction}). So, we now need to compute an average value of the probabilty to favor a packet for a given flow. We also have to take into account that, for a given size of flow $s$, only a proportion of these flows have effectively left the slow-start phase. The other ones remain in slow-start and the analysis of their probabilty to favor a packet follows the first phase.
To correctly describe this phase, we need to assess which part of flows of size $s$, $s_1 \leqslant s \leqslant s_2$, has left the slow start phase at packet $s_1$, $s_1+1$, ... $s$. As a first approximation, we use a uniform distribution between $s_1$ and $s_2$. This means that for flows of size $s$, the proportion of flows which have left the slow-start phase at $s_1$, $s_1+1$, ... $s-1$ is $\frac{1}{s_2-s_1}$ and the proportion of flows of size $s$ which have not yet left the slow-start phase is thus $\frac{s_2-s}{s_2-s_1}$.

If we denote $p_k$ the proportion of flows of size $s \geqslant s_1$ that have left the slow start-phase at $k$ we have: 

\small
\begin{eqnarray*}
& P(Favor|(S=s,Z=0)) = \\
& \sum_{i=0}^{s-s_1-1}p_k.P(Favor|k=s_1+i,Z=0,S=s)
\end{eqnarray*}
and
\begin{eqnarray*}
& P(Favor|(S=s,Z=1)) = \\
& \sum_{i=0}^{s-s_1-1}p_k.P(Favor|k=s_1+i,Z=1,S=s)
\end{eqnarray*}
\normalsize

and as in (\ref{eqn_cond1}) we obtain:

\small
\begin{eqnarray*}
& P(Favor|S=s) = \\
& P(Z=0).\sum_{i=0}^{s-s_1-1}p_k.P(Favor|k=s_1+i,Z=0,S=s)+\\ 
& P(Z=1).\sum_{i=0}^{s-s_1-1}p_k.P(Favor|k=s_1+i,Z=1,S=s)
\end{eqnarray*}
\normalsize

\subsubsection*{Third phase}

The model of this phase is quite simple. In fact, each packet of a flow which has left the slow-start phase has a probability to be favored of $70\%$. We compute the probability for a packet to be favored by taking into account the time at which a flow has left the slow-start phase and the proportion of flows as in the second phase.

\subsubsection*{Model fitting}

\begin{figure}[htb!]
   \begin{minipage}[b]{1.0\columnwidth}
	\centering
	\includegraphics[width=0.7\columnwidth]{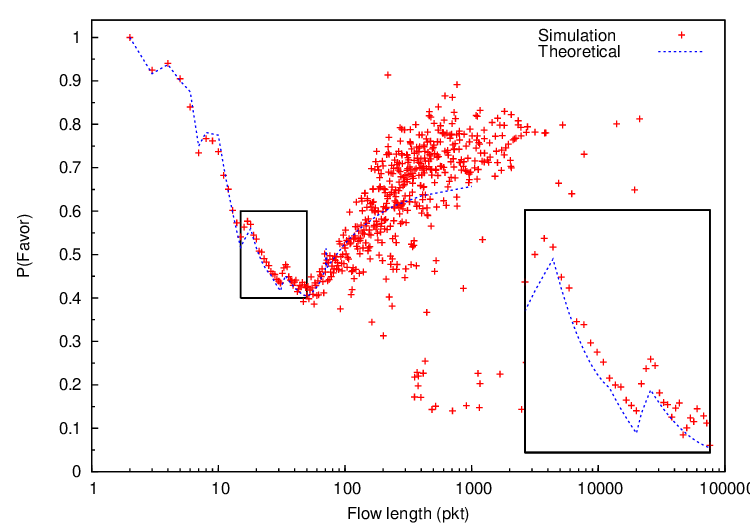}
	\caption{Model fitting for $\rho=0.25$ with $P(Z=1)=0.25$.}
	\label{fig:cond3}
   \end{minipage}
   
   \begin{minipage}[b]{1.0\columnwidth}   
	\centering
	\includegraphics[width=0.7\columnwidth]{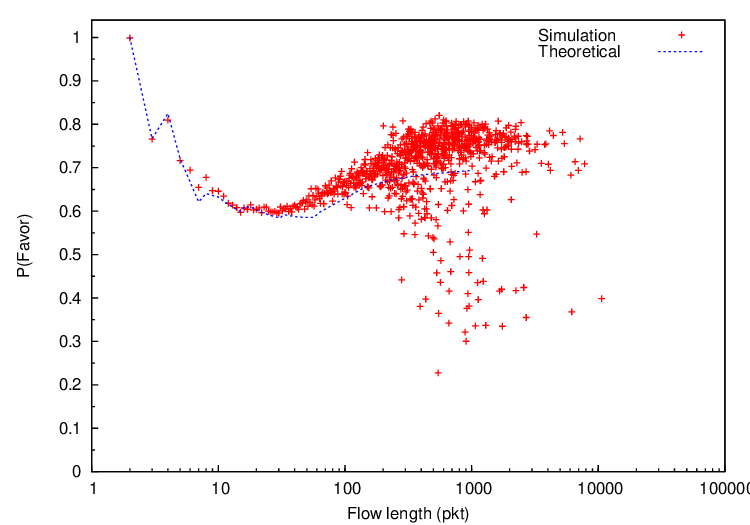}
	\caption{Model fitting for $\rho=0.85$ with $P(Z=1)=0.7$.}
	\label{fig:cond9}
   \end{minipage}
\end{figure}

To verify our model, among the ten loads that are averaged in Figure \ref{fig:probaction}, we choose two verify our model for two loads: $\rho=0.25$ and $\rho=0.85$.
For the first one we have estimated $P(Z=1)=0.25$ and $P(Z=1)=0.7$ for the second. Figures \ref{fig:cond3} and \ref{fig:cond9} show that our model correctly fits both experiments.

This model allows to understand the peaks in Figure \ref{fig:cond3} when the flow size is smaller than a hundred packets. These peaks are explained by the modelling of the first phase. Indeed, the traffic during the slow-start is bursty. Each burst has either one or two packets favored as a function of $Z$ (\textit{i.e.} up to three packets are favored when $Z=0$ and only one when $Z=1$ as given by (\ref{eq1}) and (\ref{eq2})).

\section{Deploying FavorQueue}
\label{sec:5hops}

There have been several AQM proposals from the networking community that aimed to improve the traffic flow. Although most of them have demonstrated a concrete interest for the network, a little have been effectively deployed. The deployment of a novel AQM inside the core of the network is a complex task as it requests both heavy standardisation process at the IETF and common acceptation from the router manufacturers. However, edge access routers are accessible (today, the first hop of a domestic network is usually the DSL router) and considered as the bottleneck of the network compared to the core links that follow an overprovisionning strategy \cite{poplevel}. 

\begin{figure}[htb!]
   	\centering
	\includegraphics[width=1.0\columnwidth]{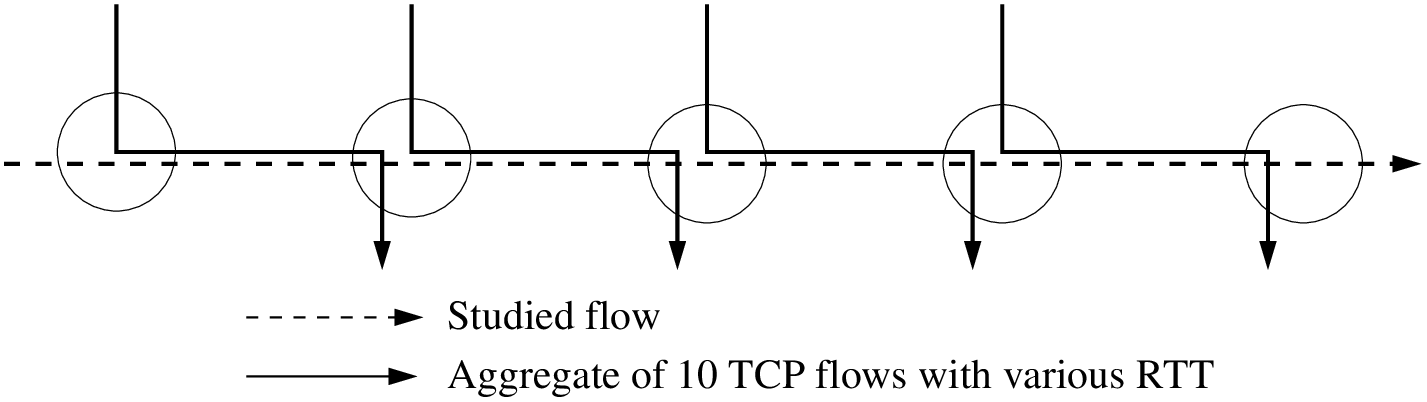}
	\caption{The five hops topology.}
	\label{fig:5hops}
\end{figure}

\begin{figure}[htb!]
   	\centering
\includegraphics[width=0.7\columnwidth]{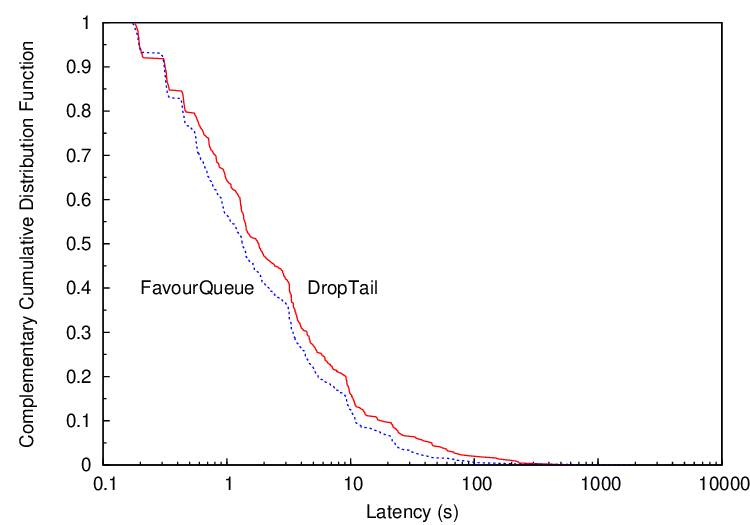}
	\caption{Comparison of the results obtained between DropTail and FavorQueue.}
	\label{fig:5hops_results}
\end{figure}

In this section, we show that a partial deployment of FavorQueue, only at the edge of the network, allows to improve the traffic performance observed by the end-user. In order to demonstrate this, we enable a five hops topology in a row as illustrated in Figure \ref{fig:5hops}. On each hop, a traffic of ten TCP long-lived flows with different RTT is generated (represented by the plain line in Figure \ref{fig:5hops}). The access link of these flows is set to 100\,Mb while the link between each hop is set to 10\,Mb where we consider a transmission delay of 1ms. The router queue size is set to 8 packets. We consider a user flow crossing the whole network (the studied flow represented by the dashed line in Figure \ref{fig:5hops}) where the access and output links have a delay set to 25\,ms. This reference flow has a finite size that follows a Pareto law of shape 1.3 and mean 30 packets. All packets have a fixed size of 1500\,B. As soon as the flow ends, a new one is triggered after a random waiting time ranging from 0 to 0.15\,sec. We run the simulation during 5000\,sec and then estimate the latency for each flow generated.
The results are given in Figure \ref{fig:5hops_results}. The DropTail curve gives the results obtained when all nodes enable a DropTail queue while the FavorQueue one shows the results when only the first hop enables FavorQueue, all the others remain with DropTail.

These results are unequivocal and show that FavorQueue enhances the performance of the end-user when only deployed at the edge. Furthermore, during the simulation, 442 flows has been generated with DropTail while FavorQueue allowed to send 538 flows. In brief, this experiment shows that the latency is improved, but this does not mean that all short flows are systematicaly favored. Indeed, the objective of this experiment is to demonstrate that a partial deployment of FavorQueue (in this case at the edge of the network) is always beneficial for the end-user. 

\section{Discussion}
\label{sec:discuss}

\subsection{Security consideration}

In the related work presented in Section \ref{sec:related}, we mention a similar solution to our proposal that gives priority to TCP packets with a SYN flag set. One of the main criticism 
that raises such kind of proposals usually deals with TCP SYN flood attack where TCP SYN packets may be used by malicious clients to improve this kind of threat \cite{rfc4987}.
However, this is a false problem as accelerating these packets does not introduce any novel security or stability side-effects as explained in \cite{Kuzmanovic05}.
Today, current TCP stacks enable protection to mitigate such well-known 
denial of service attack\footnote{See for instance \url{http://www.symantec.com/connect/articles/hardening-tcpip-stack-syn-attacks}} and current Intrusion Detection Systems (IDS) such as SNORT\footnote{\url{http://www.snort.org/}} 
combined with firewall rules, allow network providers and companies to stop such attacks. 
Indeed, the core network should not be involved in such security issue that should remain under the reponsability of edge networks and end-hosts.
Concerning the reverse path and as raised in \cite{Kuzmanovic05}, provoking web servers or hosts to send SYN/ACK packets to third parties in order to perform a SYN/ACK flood attack would be greatly inefficient. 
This is because the third parties would immediately drop such packets, since they would know that they did not generate the TCP SYN packets in the first place.

\subsection{Deployment issue}

Although there is no scalability issue anymore inside new Internet routers that can manage millions of per-flow state \cite{roberts09}. FavorQueue does not involve per-flow state management and the number of entries
that need to handle a FavorQueue router is as a function of the number of packets that can be enqueued. Furthermore, as the size of a router buffer should be small \cite{Ganjali06}, the number of states that needs to be handled is thus bounded.

To sum up, the proposed scheme respects the following constraints:
\begin{itemize}
\item easily and quickly deployable; this means that FavorQueue has no tuning parameter and does not require any protocol modification at a transport or a network level;
\item independently deployable: installation can be done without any coordination  between network operators. Operation must be done without any signaling;
\item scalable; no per-flow state is needed.
\end{itemize}

FavorQueue should be of interest for access networks; entreprise networks or universities where congestion might occur at their output Internet link.  

\subsection{About the increase of the initial slow-start value}

We wish to point out that one of the current hot topic currently discussed within the Internet Congestion Control Research Group (ICCRG) deals with the TCP initial window size. 
In a recent survey, the authors of \cite{survey10cc} highlight that the problem of short-lived flows is still not yet fully investigated and that the congestion control schemes
developed so far do not really work if the connection lifetime is only one or two RTTs. Clearly, they argue for further investigation on the impact of initial value of the congestion window on the performance of short-lived flows.
Some recent studies have also demonstrated that larger initial TCP window helps faster recovery of packet losses and as a result improves the latency in spite of increased packet losses \cite{dukkipati10}, \cite{scharf09}. 
Several proposals have also proposed solutions to mitigate the impact of the slow start \cite{rfc4782}, \cite{jumpstart}, \cite{scharf08}. 

\begin{figure}[htb!]
   	\centering
	\includegraphics[width=0.7\columnwidth]{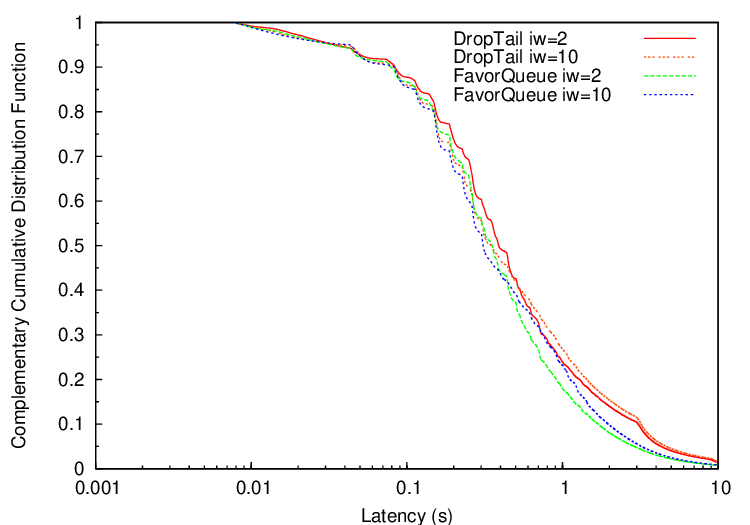}
	\caption{Comparison of the benefit obtained in terms of latency with an initial TCP window size of ten packets.}
	\label{fig:iw10latency}
\end{figure}

Although we do not act at the end-host side, we share the common goal to reduce latency during the slow start phase of a short TCP connection. However, we do not target the same objective. Indeed, end-host solutions,
that propose to increase the number of packets of the initial window, seek to mitigate the impact of the RTT loop while we seek to favor short TCP traffic when the network is congested. At the early stage of the connection, 
the number of packets exchanged is low and a short TCP request is both constrained by the RTT loop and the small amount of data exchange. Thus, some studies propose to increase this initial window value \cite{dukkipati10}, \cite{scharf09}; to change the pace at which the slow-start sends data packets by shrinking the timescale at which TCP operates \cite{kaur09rapid}; even to completely suppress the slow-start \cite{jumpstart}. Basically, all these proposals attempt 
to mitigate the impact of the slow-start loop that might be counterproductive over large bandwidth product networks. On the contrary, FavorQueue does not act on the quantity of data exchanged but prevents losses at the beginning 
of the connection.
As a result, we believe that FavorQueue must not be seen as a competitor of these end-host proposals but as a complementary mechanism. 
We propose to illustrate this complementarity by looking at the performance obtained with an initial congestion window set to ten packets. 
Figure \ref{fig:iw10latency} gives the complementary cumulative distribution function of the latency for DropTail and FavorQueue with flows with an initial slow-start set to two or ten packets. We did not change the experimental conditions (\textit{i.e.} the router buffer is still set to eight packets) and this experiment corresponds to a ten averaged experiments (see section \ref{sec:exp}). If we focus on the results obtained with DropTail for both initial window size, the increase of the initial window improves the latency (with the price of an increase of the loss rate as also denoted in \cite{dukkipati10}). However, the use of FavorQueue enforces the performance obtained and complement the action of such end-host modifications making FavorQueue a generic solution to improve short TCP traffic whatever the slow-start variant used.
Finally, Figure \ref{fig:iw10latency}  shows that FavorQueue remains compliant with recent TCP updates such as the increase of the initial slow-start value.

\section{Conclusion}
\label{sec:conclu}

In this paper, we investigate an AQM solution to accelerate short TCP flows. The main advantages of FavorQueue is to be stateless; does not require any modification inside TCP; can be used over a best effort network; does not need to be completely deployed over an Internet path. Indeed, a partial deployment could only be done over routers from an Internet service provider or over a specific AS.  

We drive several simulation scenarios showing that the drop ratio decreases for all flow lengths, thus decreasing their latency. 
FavorQueue significantly improves the performance of short TCP traffic in terms of transfer delay. The main reasons are that this mechanism 
strongly reduces the loss of a retransmitted packet triggered by an RTO and improves the connection establishment delay. 
Although FavorQueue targets short TCP flows' performance, results show that by protecting retransmitted packets, the latency of the whole traffic and particularly non-opportunistic flows, is improved. 

In a future work, we aim at investigating FavorQueue with rate-based transport protocols such as TFRC in order to verify whether we would benefit similar properties and with delay-based TCP protocol variants (such as TCP Vegas and TCP Compound) that should intuitively take large benefit of such AQM. We also expect to enable ECN support in FavorQueue.

\section*{Acknowledgements}
The authors wish to thank Eugen Dedu for numerous discussions about this work and Olivier Mehani for helpful comments. 

\bibliographystyle{unsrt}
\bibliography{biblio}

\begin{thebibliography}{10}

\bibitem{IOR2009}
C.~Labovitz, S.~Iekel-Johnson, D.~McPherson, J.~Oberheide, F.~Jahanian, and
  M.~Karir.
\newblock Atlas internet observatory 2009 annual report.
\newblock In {\em 47th NANOG}, 2009.

\bibitem{ciullo09}
D.~Ciullo, M.~Mellia, and M.~Meo.
\newblock Two schemes to reduce latency in short lived {TCP} flows.
\newblock {\em Communications Letters}, 13(10), October 2009.

\bibitem{chen03}
X.~Chen and J.~Heidemann.
\newblock Preferential treatment for short flows to reduce web latency.
\newblock {\em Computer Networks}, 41(6):779--794, April 2003.

\bibitem{rfc1122}
R.~Braden.
\newblock Requirements for internet hosts.
\newblock RFC 1122, October 1989.

\bibitem{Kantawala02}
A.~Kantawala and J.~Turner.
\newblock Queue management for short-lived {TCP} flows in backbone routers.
\newblock In {\em GLOBECOM}, pages 2380--2384. IEEE, November 2002.

\bibitem{avrachenkov04}
K.~Avrachenkov, U.~Ayesta, P.~Brown, and E.~Nyberg.
\newblock Differentiation between short and long {TCP} flows: Predictability of
  the response time.
\newblock In {\em INFOCOM}, Hong Kong, March 2004. IEEE.

\bibitem{rai05}
I.A. Rai, E.W. Biersack, and G.~Urvoy-Keller.
\newblock Size-based scheduling to improve the performance of short {TCP}
  flows.
\newblock {\em Network}, 19(1):12--17, January 2005.

\bibitem{Kleinrock75}
L.~Kleinrock.
\newblock {\em Queueing systems}, volume~1 of {\em Wiley Interscience}.
\newblock John Wiley \& Sons, 1975.

\bibitem{pan00}
R.~Pan, B.~Prabhakar, and K.~Psounis.
\newblock Choke: A stateless {AQM} scheme for approximating fair bandwidth
  allocation.
\newblock In {\em INFOCOM}. IEEE, 2000.

\bibitem{mellia02}
M.~Mellia, I.~Stoica, and H.~Zhang.
\newblock {TCP}-aware packet marking in networks with diffserv support.
\newblock {\em Computer Networks}, 42(1):81--100, 2003.

\bibitem{Guo01}
L.~Guo and L.I. Matta.
\newblock The war between mice and elephants.
\newblock In {\em ICNP}, Riverside, California, November 2001. IEEE.

\bibitem{rfc2475}
S.~Blake, D.~Black, M.~Carlson, E.~Davies, Z.~Wang, and W.~Weiss.
\newblock An architecture for differentiated service.
\newblock RFC 2475, December 1998.

\bibitem{rfc5562}
A.~Kuzmanovic, A.~Mondal, S.~Floyd, and K.K. Ramakrishnan.
\newblock Adding {E}xplicit {C}ongestion {N}otification ({ECN}) capability to
  {TCP}'s {SYN/ACK} packets.
\newblock RFC 5562, June 2009.

\bibitem{dedu09pdp}
E.~Dedu and E.~Lochin.
\newblock A study on the benefit of {TCP} packet prioritisation.
\newblock In {\em PDP}, Weimar, Germany, February 2009.

\bibitem{Appenzeller04}
Guido Appenzeller, Isaac Keslassy, and Nick McKeown.
\newblock Sizing router buffers.
\newblock In {\em SIGCOMM}, pages 281--292. ACM, 2004.

\bibitem{Arthur07}
C.M. Arthur, A.~Lehane, and D.~Harle.
\newblock Keeping order: Determining the effect of {TCP} packet reordering.
\newblock In {\em ICNS}, Athens, Greece, June 2007.

\bibitem{Lachlan2008}
L.~Andrew, C.~Marcondes, S.~Floyd, L.~Dunn, R.~Guillier, G.~Wang, L.~Eggert,
  S.~Ha, and I.~Rhee.
\newblock Towards a common {TCP} evaluation suite.
\newblock In {\em Workshop on Protocols for Fast Long-Distance Networks
  (PFLDnet)}, Manchester, March 2008.

\bibitem{Ganjali06}
Y.~Ganjali and N.~McKeown.
\newblock Update on buffer sizing in internet routers.
\newblock {\em Computer Communication Review}, 36(5), October 2006.

\bibitem{Barakat00}
C.~Barakat and E.~Altman.
\newblock Performance of short {TCP} transfers.
\newblock In {\em NETWORKING}, Paris, May 2000. Springer-Verlag.

\bibitem{mathis97macroscopic}
M.~Mathis, J.~Semke, and J.~Mahdavi.
\newblock The macroscopic behavior of the {TCP} congestion avoidance algorithm.
\newblock {\em Computer Communications Review}, 27(3), 1997.

\bibitem{poplevel}
Supratik Bhattacharyya, Christophe Diot, and Jorjeta Jetcheva.
\newblock Pop-level and access-link-level traffic dynamics in a tier-1 pop.
\newblock In {\em Proceedings of the 1st ACM SIGCOMM Workshop on Internet
  Measurement}, pages 39--53, New York, NY, USA, 2001. ACM.

\bibitem{rfc4987}
W.~Eddy.
\newblock {{TCP SYN} Flooding Attacks and Common Mitigations}.
\newblock RFC 4987 (Informational), August 2007.

\bibitem{Kuzmanovic05}
A.~Kuzmanovic.
\newblock The power of explicit congestion notification.
\newblock In {\em SIGCOMM}, pages 61--72, Philadelphia, August 2005. ACM.

\bibitem{roberts09}
L.~Roberts.
\newblock A radical new router.
\newblock {\em Spectrum}, 46(7), July 2009.

\bibitem{survey10cc}
A.~Afanasyev, N.~Tilley, P.~Reiher, and L.~Kleinrock.
\newblock Host-to-host congestion control for {TCP}.
\newblock {\em Communications Surveys \& Tutorials}, 12(3):304 --342, 2010.

\bibitem{dukkipati10}
N.~Dukkipati et~al.
\newblock An argument for increasing {TCP}'s initial congestion window.
\newblock {\em Computer Communication Review}, 40(3), July 2010.

\bibitem{scharf09}
M.~Scharf.
\newblock Performance evaluation of fast startup congestion control schemes.
\newblock In {\em NETWORKING}, Aachen, Germany, May 2009. Springer-Verlag.

\bibitem{rfc4782}
S.~Floyd, M.~Allman, A.~Jain, and P.~Sarolahti.
\newblock {Quick-Start for {TCP} and {IP}}.
\newblock RFC 4782 (Experimental), January 2007.

\bibitem{jumpstart}
Dan Liu, Mark Allman, Shudong Jin, and Limin Wang.
\newblock Congestion control without a startup phase.
\newblock In {\em Workshop on Protocols for Fast Long-Distance Networks
  (PFLDnet)}, 2007.

\bibitem{scharf08}
Michael Scharf and Haiko Strotbek.
\newblock Performance evaluation of quick-start {TCP} with a {Linux} kernel
  implementation.
\newblock In {\em NETWORKING}, 2008.

\bibitem{kaur09rapid}
V.~Konda and J.~Kaur.
\newblock {RAPID}: Shrinking the congestion-control timescale.
\newblock In {\em INFOCOM}. IEEE, April 2009.

\end{thebibliography}
\end{document}